\documentstyle[12pt]{article}

\def\pd{\partial}
\def\a{\alpha}
\def\b{\beta}
\def\dl{\delta}
\def\s{\sigma}

\def\vphi{\varphi}
\def\eps{\epsilon}
\def\veps{\varepsilon}
\def\lam{\lambda}
\def\Lam{\Lambda}
\def\hg{{\hat g}}
\def\hnabla{{\hat \nabla}}
\def\hR{{\hat R}}

\def\hE{{\hat E}}
\def\hDelta{{\hat \Delta}}
\def\bmu{{\bar \mu}}
\def\bnu{{\bar \nu}}
\def\blam{{\bar \lambda}}
\def\bs{{\bar \sigma}}
\def\prm{m^{\prime}}
\def\prn{n^{\prime}}
\def\prs{s^{\prime}}
\def\prt{t^{\prime}}
\def\pru{u^{\prime}}

\def\prp{p^{\prime}}
\def\ta{\tilde{a}}
\def\tb{\tilde{b}}
\def\tc{\tilde{c}}
\def\td{\tilde{d}}
\def\te{\tilde{e}}
\def\tq{\tilde{q}}
\def\tvphi{\tilde{\varphi}}

\def\gm{\gamma}
\def\hgm{{\hat \gamma}}
\def\Gm{\Gamma}
\def\om{\omega}
\def\Om{\Omega}
\def\sq{\sqrt}
\def\e{\hbox{\large \it e}}
\def\half{\frac{1}{2}}
\def\fr{\frac}
\def\pp{\prime}
\def\arr{\rightarrow}
\def\C{{\bf C}}
\def\D{{\bf D}}
\def\tD{\tilde{{\bf D}}}
\def\E{{\bf E}}
\def\G{{\bf G}}
\def\H{{\bf H}}
\def\I{{\bf I}}
\def\bM{\overline{M}}
\def\uM{\underline{M}}
\def\bb{\begin{equation}}
\def\ee{\end{equation}}
\def\bba{\begin{eqnarray}}
\def\eea{\end{eqnarray}}

\begin{document}

\begin{titlepage}

\begin{tabbing}
   qqqqqqqqqqqqqqqqqqqqqqqqqqqqqqqqqqqqqqqqqqqqqq 
   \= qqqqqqqqqqqqq  \kill 
         \>  {\sc KEK-TH-900}    \\
         \>       hep-th/0307008 \\
         \>  {\sc July 2003} 
\end{tabbing}

\begin{center}
{\Large {\bf Conformal Algebra and Physical States  
in a Non-Critical 3-Brane on $R \times S^3$}}
\end{center}

\centering{\sc Ken-ji Hamada\footnote{E-mail address : 
hamada@post.kek.jp} 
and Shinichi Horata\footnote{E-mail address : 
horata@post.kek.jp}}

\begin{center}
{\it Institute of Particle and Nuclear Studies, \break 
High Energy Accelerator Research Organization (KEK),} \\ 
{\it Tsukuba, Ibaraki 305-0801, Japan}
\end{center}

\begin{abstract} 
A world-volume model of a non-critical 3-brane is quantized in a strong coupling 
phase in which fluctuations of the conformal mode become dominant. 
This phase, called the conformal-mode dominant phase,
is realized at very high energies far beyond the Planck mass scale. 
We separately treat the conformal mode and the traceless mode    
and quantize the conformal mode {\it non-perturbatively}, while the traceless mode is 
treated in a perturbative method that is renormalizable and asymptotically free.  
In the conformal-mode dominant phase, the coupling of the traceless mode vanishes, 
and the world-volume dynamics are described by a four-dimensional conformal 
field theory (${\rm CFT}_4$). We canonically quantize this 
model on $R \times S^3$, where the dynamical fields are expanded 
in spherical tensor harmonics on $S^3$, which include both positive-metric 
and negative-metric modes.   
Conformal charges and a conformal algebra are constructed.  
They yield strong constraints on physical states.  
We find that all negative-metric modes are related to positive-metric modes 
through the charges, and thus negative-metric modes are themselves {\it not} independent 
physical modes. Physical states  satisfying the conformal invariance conditions 
are given by particular combinations of positive-metric and negative-metric modes. 
An infinite number of such physical states are constructed. 
In the appendices, we construct spherical vector and tensor harmonics 
on $S^3$ in practical forms using the Wigner $D$ functions and the Clebsch-Gordan 
coefficients and calculate the integrals of three and four products of these harmonics 
over $S^3$. 


\end{abstract}
\end{titlepage}  

\section{Introduction}
\setcounter{equation}{0}

\noindent

  In this paper, we quantize a world-volume model of a 3-brane.  
As in the case of non-critical 
string~\cite{polyakov, kpz, dk, bk, seiberg, bmp, hamada93, kkn, klebanov},  
the action for the conformal mode of the world-volume metric is induced from 
the measure as the Wess-Zumino action~\cite{wz} (Liouville action~\cite{polyakov} 
in two dimensions) related to the conformal anomaly~\cite{cd, ddi, duff, fujikawa, bcr}.  
The conformal mode is treated {\it non-perturbatively} in the sense that there is 
no coupling concerning this mode. 
In four dimensions, we cannot make the conformal anomaly vanish.  
Fo this reason, we refer to a quantum theory of a 3-brane as a non-critical 
3-brane~\cite{riegert, ft84, am, amm92, amm97, hs, hamada02}. 
Furthermore, considering a world-volume model coupled to 
four-dimensional ${\cal N}=4$ super Yang-Mills theory (${\rm SYM}_4$), 
it is a world-volume model of a D3-brane~\cite{polchinski, nst}.    
The aim of this paper is to quantize a non-critical 3-brane on the $R\times S^3$ 
background, derive the conformal algebra corresponding 
to the Virasoro algebra in a non-critical string on $R\times S$, and construct 
physical states.

  Of course, in four dimensions, the metric field becomes dynamical, 
and we must solve the problem of world-volume singularities in the strong coupling phase. 
We here consider a model in which singularities are removed quantum mechanically. 
In this sense, the Einstein theory is excluded, because singularities are 
solutions of $R_{\mu\nu}=0$, so that the quantum weight of the Einstein action 
becomes order 1: $\e^{-\int R} \sim o(1)$. 
To begin with, the action is not bounded from below.

  We here look for a model in which the world-volume becomes conformally flat in the 
strong coupling phase of gravity realized in the energy region, 
i.e. that in which $E \gg M_P$, 
where $M_P$ is the Planck mass~\cite{hs, hamada02}.  
We call this strong coupling phase the {\it conformal-mode dominant phase}.
This implies that configurations in which 
the Weyl tensor $C_{\mu\nu\lam\s}$ vanishes become dominant in this phase.

  The conformal-mode dominant phase is realized by the quantum 
weight $\exp (-\fr{1}{t^2}\int C_{\mu\nu\lam\s}^2 )$~\cite{stelle, tomboulis, ft82, 
bhs, bg, hs, hamada02}.  
This weight implies that singular configurations 
have vanishing weights,~\footnote{ 
The singularity that satisfies $C_{\mu\nu\lam\s}=0$ is not considered here.  
However, this singularity is removed by the action of the conformal mode induced 
from the measure.} 
and the $C_{\mu\nu\lam\s}=0$ configuration 
becomes dominant in the limit $t \arr 0$. Here, $t$ is the coupling of the 
traceless mode. Thus, the asymptotic freedom~\cite{ft82, hamada02} now implies 
that the world-volume becomes conformally flat at very high energies far 
beyond the Planck scale.  
This feature is called an {\it asymptotically conformal flatness} ~\cite{hs,hamada02}.    
Here, it is worth commenting that this weight gives the same value for 
all configurations for which only the values of the conformal mode are 
different, because of the conformal invariance of the Weyl action. 
Thus, we must integrate the conformal mode exactly over the entire range.

  The fact that there is no singularity implies that point-like excitations are forbidden 
for $E \geq M_P$, because for such energies, a particle is a mini black hole.  
This is an advantageous feature of the model when we consider unitarity 
in the strong coupling phase, i.e. the information loss problem.  
In fact, at such energies, the Heisenberg 
uncertainty for the position, $\Delta x \sim \fr{\hbar }{M_P c}=L_P$, becomes smaller than 
the Schwarzshild radius, $\fr{2G_N M_P}{c^2}=2L_P$, where $G_N$ is the 
gravitational constant and $L_P$ is the Planck length.    
Thus, all information regarding such a point-like excitation is inside the black hole, 
so that it is lost, because we cannot extract it. 
This means that we must abandon the standard particle picture for $E \geq M_P$. 
The fact that such singularities are forbidden quantum mechanically 
implies that physical states are changed in this energy region.

  The conformal invariance determines the kind of world-volume excitations that arise 
in the conformal-mode dominant phase. This is related to the problem of negative-metric 
modes in higher-derivative actions.  Because we cannot remove such undesired modes 
using gauge degrees of freedom, we need a {\it new mechanism} to confine  
them~\cite{tomboulis, bhs, bg}. 
We here assert that the conformal invariance is just such a new mechanism. 
In order to solve this problem, we construct conformal charges and a conformal algebra 
that define physical states in the strong coupling phase. 
In fact, we find that {\it conformal charges relate negative-metric modes to 
positive-metric modes so that no negative-metric modes are independent.   
Physical states satisfying the conformal invariance condition are given by 
particular combinations of positive-metric and negative-metric modes}.  
There are an infinite number of such physical states. 
We construct them in definite forms.

  This paper is organized as follows. In the next section, we define a model of 
a non-critical 3-brane and summarize the advantageous properties of this model 
and the problems we must solve. 
In Sect.3 we canonically quantize the model on $R \times S^3$ 
in the conformal-mode dominant phase. 
We here introduce the ${\it radiation}^+$ gauge, in which the space of the residual 
gauge symmetry becomes equivalent to the space spanned by the conformal Killing vectors 
on $R\times S^3$. The mode expansions and canonical commutation relations are 
listed there. In Sect.4, we construct conformal charges and their algebra for all dynamical 
fields. The scalar fields and the conformal mode have already been constructed 
by Antoniadis, Mazur and Mottola~\cite{amm97}. We develop their calculations, as we can   
treat fields with vector and tensor indices, in particular the traceless mode. 
In Sect.5,  physical state conditions are defined. 
There, we give definite expressions of physical states.    
The last section is devoted to conclusions. 
In the appendices we develop the technique for spherical tensor harmonics on $S^3$. 
We construct them in practical forms using the Wigner $D$ functions and the Clebsch-Gordan 
coefficients and calculate the integrals of products of three and four harmonics. 
Useful formulae, product expansions, crossing relations, and so on, are obtained there.

\section{A World-volume Model of a Non-critical 3-brane}
\setcounter{equation}{0}
\noindent

  We would like to construct a non-critical 3-brane theory without world-volume  singularities. 
We here propose a non-singular model in which the world-volume metric becomes conformally flat 
in the strong coupling phase of gravity. 
This implies that fluctuations of the conformal mode become dominant 
in the region of energies much higher than the Planck energy. 
Such a configuration is characterized by the vanishing of the Weyl tensor.

  A model satisfying this  non-singular condition was proposed 
in Refs.\cite{hs, hamada02}. This model is
defined, in the Lorentzian signature $(-1,1,1,1)$, by the partition 
function
\bb
  Z=\int [g^{-1}dg]_g [dA]_g [dX]_g \exp (iI), 
   \label{Z}
\ee
with the action  
\bba
  && I=\int d^4 x \sq{-g} \biggl\{ -\fr{1}{t^2} C_{\mu\nu\lam\s}C^{\mu\nu\lam\s} 
     -b G_4 +\fr{M_G^2}{2} R -\Lam \\
        \nonumber
  && \qquad\qquad\qquad\qquad
    -\fr{1}{4}F_{\mu\nu}F^{\mu\nu} 
    - \fr{1}{2} \left( \pd^{\mu} X \pd_{\mu}X +\fr{1}{6}R X^2 \right) \biggr\}. 
\eea
Here $M_G=M_P/\sq{8\pi}$ is the reduced Planck mass. 
$t$ is the coupling constant of the traceless mode, and $C_{\mu\nu\lam\s}$ 
is the Weyl tensor,
\bb
  C_{\mu\nu\lam\s} = R_{\mu\nu\lam\s} 
     - \half ( g_{\mu\lam}R_{\nu\s}-g_{\mu\s}R_{\nu\lam}
                   -g_{\nu\lam}R_{\mu\s}+g_{\nu\s}R_{\mu\lam} ) 
     +\fr{1}{6} ( g_{\mu\lam}g_{\nu\s}-g_{\mu\s}g_{\nu\lam} ) R,        
\ee
which is the field strength of the traceless mode. Also, $G_4$ is the Euler 
density in 4 dimensions,
\bba
    G_4 &=& \fr{1}{4}\eps^{\mu\nu\a\b}\eps^{\lam\s\gm\dl} 
              R_{\mu\nu\gm\dl}R_{\lam\s\a\b} 
                \nonumber  \\ 
        &=& R^{\mu\nu\lam\s}R_{\mu\nu\lam\s}-4 R^{\mu\nu}R_{\mu\nu}+R^2.  
\eea 
The bare constant $b$ is not an independent coupling. The renormalization procedure 
for this constant is given in Ref.\cite{hamada02}.
The fields $X$ and $A$ are the scalar and gauge fields. 
Considering a model coupled to ${\cal N}=4$ ${\rm SYM}_4$~\cite{nst}, 
it can be regarded as a world-volume model of a D3-brane~\cite{polchinski}.

  We here summarize the advantageous features of this world-volume model:
\begin{itemize} 
\item The perturbative expansion in the coupling, $t$, is renormalizable 
at higher order~\cite{hamada02} and asymptotically free~\cite{ft82}. 
This implies that at very high energies much greater than the Planck scale, 
configurations in which the Weyl tensor vanishes become dominant.  
This condition is refered to as 
{\it asymptotically conformal flatness} (ACF)~\cite{hs,hamada02}. 
\item ACF implies that fluctuations of the conformal mode become important.  
This corresponds to a strong coupling phase of gravity, 
called  the {\it conformal-mode dominant phase}. 
In this phase, the conformal mode must 
be quantized non-perturbatively, so that the model can be described as ${\rm CFT}_4$.  
\item There is no pure $R^2$ action, because it is forbidden by the condition of 
integrability,\cite{bcr, riegert, ft84}\footnote{ 
The integrability condition is expressed by 
the equation $[\dl_{\om_1}, \dl_{\om_2}] \Gamma =0$, where $\dl_{\om}$ is 
a conformal variation and $\Gamma$ is the 1PI gravitational effective 
action~\cite{bcr}. This is the condition determining whether $\Gamma$ exists, 
because $\Gamma$ is obtained by integrating the ordinary action with respect to 
the conformal mode. This condition, in addition to forbidding the $R^2$ action,   
requires the action to be a combination of $C_{\mu\nu\lam\s}^2$ and $G_4$    
and also requires conformal coupling for scalar fields. 
On the other hand, there is no restriction for actions with dimensional parameters, 
such as the Einstein action, the cosmological constant and mass terms of matter fields.  
} 
or higher-order renormalizability~\cite{hamada02}. 
The kinetic term of the conformal mode is induced from the measure, 
as in the case of a non-critical string. This fact is also important to ensure ACF. 
\item ACF implies that there is {\it no} singular excitation in the strong coupling 
phase.  This implies that physical states change in this phase.   
\end{itemize}

  The world-volume metric in the conformal-mode dominant phase is 
given by~\cite{kkn,hs,hamada02} 
\bb
      g_{\mu\nu} =\e^{2\phi} \hg_{\mu\lam} 
                   (\dl^{\lam}_{~\nu} +t h^{\lam}_{~\nu} + \cdots ), 
       \qquad  tr(h)=0,      
\ee
where $h^{\mu}_{~\nu}$ is the traceless mode.  
Note that there is {\it no} coupling for the conformal mode, $\phi$.

  Of course, there is a strong coupling phase in which the traceless-mode coupling 
becomes strong. We call this phase the {\it traceless-mode dominant phase}. It is 
realized near the Planck energy, where we must take into account the ordinary strong 
coupling ingredients, like gravitational instantons~\cite{shp}.  
In the energy region much below the Planck scale, the Einstein action becomes dominant. 
This is the classical phase, in which the standard weak-field approximation becomes valid.

  The kinetic term and the interaction terms of the conformal mode are induced 
from the measure~\cite{hamada02}.  
In the limit of the vanishing  coupling $t$, only the kinetic term remains, 
and the partition function (\ref{Z}) can be equivalently re-expressed as
\bb
    Z(\hg)=\int [d\phi]_{\hg}[dh]_{\hg}[dX]_{\hg}[dA]_{\hg}\exp ( iI_{\rm CFT} ), 
\ee
where $I_{\rm CFT}=S+I_{t=0}$, and $S$ is the induced  Wess-Zumino action 
(the local part of the Riegert action~\cite{riegert}) 
related to the conformal anomaly for $G_4$,  
\bb
   S=-\fr{b_1}{(4\pi)^2} \int d^4 x \sq{-\hg} \left\{ 
       2 \phi \hDelta_4 \phi + \hE_4 \phi \right\} .
\ee
Here, $\sq{-g}\Delta_4$ is the conformally invariant 4-th order operator given by 
\bb
    \Delta_4 = \Box^2 +2 R^{\mu\nu}\nabla_{\mu}\nabla_{\nu}-\fr{2}{3}R\Box 
                  + \fr{1}{3} (\nabla^{\mu}R)\nabla_{\mu}, 
\ee
and $E_4 = G_4 -\fr{2}{3}\Box R$, 
which satisfies the equation $\sq{-g}E_4=\sq{-\hg}(4\hDelta_4 \phi +\hE_4)$ concerning 
the metric $g_{\mu\nu}= \e^{2\phi}\hg_{\mu\nu}$. 
This action is the four-dimensional counteraction of the Liouville action 
(the local part of the Polyakov action~\cite{polyakov}) for a non-critical 
string. The coefficient $b_1$ has been calculated as
\bb
    b_1=\fr{1}{360}(N_X +62 N_A) + \fr{769}{180},
       \label{b-1}
\ee 
where $N_X$ and $N_A$ are the number of scalar fields and gauge fields~\cite{duff}. 
The last term in (\ref{b-1}) is the sum of $87/20$~\cite{ft82} and $-7/90$~\cite{amm92}, 
which are the contributions from the traceless mode and conformal mode, respectively.  
The conformal anomaly coefficients related to  $C_{\mu\nu\lam\s}^2$ 
and $F_{\mu\nu}F^{\mu\nu}$, which are related to $\b$-functions, depend on the 
coupling. Therefore they vanish in the limit of vanishing coupling.

  It is worth commenting that the conformal anomaly is physically {\it not} an anomalous 
quantity.  It is a necessary ingredient to preserve diffeomorphism invariance. 
Now, the metric is a dynamical variable, so that conformal invariance 
is a part of the diffeomorphism invariance/background-metric independence. 
Thus, the conformal anomaly is a necessary ingredient to preserve conformal 
invariance itself.

  Lastly, we summarize a problem that we must solve, which is the purpose of this paper. 
It is well known that higher-derivative actions introduce 
undesired negative-metric modes, which cannot be 
removed by gauge symmetry. In general, such modes must be removed. 
However, the negative-metric modes in the metric fields are necessary to  
remove a world-volume singularity. 
It is necessary that there exist no singularity in order to 
resolve the unitarity problem in the strong coupling phase,  
called the `information loss problem'. 
Therefore, we think that such undesired modes are actually necessary to 
formulate a world-volume theory of a non-critical 3-brane, and 
we conjecture that there is a yet unrecognized mechanism to confine them. 
We here assert that this mechanism is {\it conformal invariance}.

\section{Canonical Quantization of a Non-critical 3-brane on $R \times S^3$}
\setcounter{equation}{0}
\noindent
 
  Quantum conformal invariance in the conformal-mode 
dominant phase is now manifest, because we integrate the conformal mode exactly 
over its entire domain without introducing the coupling concerning this mode.  
Thus, the partition function 
\bb
    Z(\hg)=\int^{\infty}_{-\infty} d\phi {\cal Z}(\e^{2\phi}\hg) 
\ee
satisfies the equation 
\bb
   Z(\e^{2\om}\hg)=\int^{\infty}_{-\infty} d\phi {\cal Z}(\e^{2(\phi+\om)}\hg) 
      =\int^{\infty}_{-\infty} d\phi^{\pp} {\cal Z}(\e^{2\phi^{\pp}}\hg)
      =Z(\hg) 
      \label{conf-inv}
\ee
for any local function $\om$; that is, any conformal change of the 
background-metric can be absorbed through a local shift of the conformal mode.  
Thus, a non-critical 3-brane in the conformal-mode dominant phase is 
described by ${\rm CFT}_4$. (For recent developments of ${\rm CFT}_4$ see, for example, 
Refs.\cite{maldacena, cardy, fp}).

  To quantize the model, we must specify the background metric, $\hg_{\mu\nu}$. 
Owing to the conformal invariance, all models transformed by the conformal 
transformation into each other are equivalent.  
We here choose the $R \times S^3$ metric, given by
\bba 
     d{\hat s}^2_{R\times S^3} &=&-dt^2 + \hgm_{ij}dx^i dx^j 
            \nonumber  \\ 
                 &=&-dt^2 + \fr{1}{4} (d\a^2 +d\b^2 +d\gm^2 +2 \cos \b d\a d\gm ), 
\eea
where $x^i=(\a,\b,\gm)$, with $i=1,2,3$, are the Euler angles and $t$ is the time.  
Henceforth, we mainly use $t$ to denote the time.   
The curvatures are given by $\hR_{0\mu\nu\lam}=\hR_{0\mu}=0$, 
$\hR_{ijkl}=(\hgm_{ik}\hgm_{jl}-\hgm_{il}\hgm_{jk})$, $\hR_{ij}=2\hgm_{ij}$ and $\hR=6$. 
Here, we take radius of $S^3$ to be unity. Thus, the missing dimensions in the expressions 
given below originate from the radius.  
The advantages of this metric are that mode expansions and canonical commutation relations  
have simple, diagonal forms, and the level of states is parametrized 
by a single number, $J$,  
as discussed below.\footnote{
In a flat background, $M_4$ or $R \times T^3$, the mode expansion of a 4-th order 
field is given by~\cite{pu}  
$$
    \phi =\half \int \fr{d^3 {\bf p}}{(2\pi)^{3/2}} \fr{1}{|{\bf p}|^{3/2}} 
       \left[ \left\{ a({\bf p}) +i t b({\bf p}) \right\} \e^{ip_{\mu} x^{\mu}} 
        + {\rm h.c.} \right] ,  
$$
where $p^{\mu}=(p^0, {\bf p})$. 
There is an unusual time dependence in the undesired mode, $b$, 
which does not arise in the case of $R\times S^3$.  
The commutation relations have the off-diagonal forms  
$ [ a({\bf p}), a^{\dag}({\bf q}) ]=\dl^3({\bf p}-{\bf q})$,  
$ [ a({\bf p}), b^{\dag}({\bf q}) ] = [ b({\bf p}), a^{\dag}({\bf q}) ] 
    = |{\bf p}| \dl^3({\bf p}-{\bf q}) $ 
and $ [ b({\bf p}), b^{\dag}({\bf q}) ] =0 $.         
}

\subsection{$\hbox{Radiation}^+$ gauge}
\noindent

  The kinetic term of the traceless mode is invariant under the gauge transformation 
defined by $\dl_{\xi}h_{\mu\nu}=\hnabla_{\mu}\xi_{\nu}+\hnabla_{\nu}\xi_{\mu}
-\fr{1}{2}\hg_{\mu\nu}\hnabla^{\lam}\xi_{\lam}$. 
To quantize the traceless mode, let us first consider the radiation gauge, in which
\bb
    h^0_{~0}=\hnabla_i h^i_{~0}=\hnabla_i h^i_{~j}=0. 
      \label{gauge}
\ee 
We also use the radiation gauge for the vector field, so that $A^0=\hnabla_i A^i=0$.
The combined action of $S$ and $I$ on $R \times S^3$ in the radiation gauge is 
given by 
\bba
    I_{\rm CFT} &=& \int dt \int_{S^3} d^3 x \sq{\hgm} \biggl\{ 
     -\fr{2b_1}{(4\pi)^2}\phi \left( 
          \pd_t^4 -2\hnabla^2\pd_t^2 +\hnabla^4 + 4\pd_t^2 \right) \phi 
                 \nonumber  \\ 
     && \qquad\qquad
        -\fr{1}{2}h^i_{~j} \left( \pd_t^4 -2\hnabla^2\pd_t^2 + \hnabla^4 
                     + 8\pd_t^2-4\hnabla^2 +4 \right) h^j_{~i} 
                 \nonumber    \\
     && \qquad\qquad
       + h^0_{~i} \left( \hnabla^2+2 \right) 
          \left( -\pd_t^2 +\hnabla^2 -2 \right) h^{0i}  
                 \nonumber \\ 
     && \qquad\qquad
       +\fr{1}{2}X \left( -\pd_t^2+\hnabla^2-1 \right) X 
                  \nonumber  \\ 
     && \qquad\qquad
       +\fr{1}{2}A_i \left( -\pd_t^2 +\hnabla^2 -2 \right) A^i \biggr\},                           
\eea
where $\hnabla^2=\hnabla^i \hnabla_i$ is the Laplacian on $S^3$.

  In the radiation gauge, the fields can be expanded in spherical scalar, vector and 
tensor harmonics on $S^3$, which are the eigenfunctions of the Laplacian. 
Because the isometry group on $S^3$ is $SO(4)=SU(2)\times SU(2)$, they can be 
classified using the $(J_L,J_R)$ representations 
of $SU(2)\times SU(2)$, where $J_L$ and $J_R$ take integer or half-integer values.

  The scalar harmonics, denoted by $Y_{JM}$, with $J \geq 0$, 
belong to the $(J,J)$ representation 
of $SU(2)\times SU(2)$, which satisfies the eigenequation 
\bb
     \hnabla^2 Y_{JM} =-2J(2J+2) Y_{JM},   
\ee
where $M=(m,\prm)$ and $m, \prm=-J, \cdots,J$. The multiplicity of $Y_{JM}$ is 
given by the product of the left and the right $SU(2)$ multiplicities, so that it 
is totally $(2J+1)^2$.

  The transverse vector harmonics, denoted by $Y^i_{J(My)}$, 
with $J \geq \fr{1}{2}$ and $y=\pm\half$, 
belong to the $(J+y, J-y)$ representation. They satisfy the eigenequation 
\bb
     \hnabla^2 Y^i_{J(My)} = \{ -2J(2J+2)+1 \} Y^i_{J(My)},   
         \label{eigen-v}
\ee
where $M=(m,\prm)$ and 
\bba
    m &=&-J-y,~-J-y+1, \cdots , J+y-1,~J+y, 
           \nonumber \\
    \prm &=& -J+y,~-J+y+1, \cdots , J-y-1,~J-y .
\eea
Thus, the multiplicity is  $2J(2J+2)$ for each sign of $y$, 
so that it is totally $2(2J)(2J+2)$.

  The symmetric transverse traceless tensor harmonics of rank 2, denoted 
by $Y^{ij}_{J(Mx)}$, with  $J \geq 1$ and $x=\pm 1$, belong to 
the $(J+x, J-x)$ representation, and hence satisfy the eigenequation 
\bb
     \hnabla^2 Y^{ij}_{J(Mx)} = \{ -2J(2J+2)+2 \} Y^{ij}_{J(Mx)},   
\ee
where $M=(m,\prm)$ and 
\bba
    m &=&-J-x,~-J-x+1, \cdots , J+x-1,~J+x, 
           \nonumber \\
    \prm &=& -J+x,~-J+x+1, \cdots , J-x-1,~J-x .
\eea
Thus, the multiplicity is given by  $(2J-1)(2J+3)$ for each sign of $x$, 
so that it is totally $2(2J-1)(2J+3)$.

  We wish to make the space of residual gauge symmetry equivalent to the space spanned by  
the conformal Killing vectors on $R \times S^3$. The residual symmetry space 
in the radiation gauge (\ref{gauge}) defined by the equations 
$\dl_{\xi}h^0_{~0}= \dl_{\xi} (\hnabla_i h^i_{~0}) = \dl_{\xi} (\hnabla_i h^i_{~j})=0 $  
includes the vector $\xi^{\mu}=\left( 0, f(t)Y^i_{1/2 (My)} \right)$, 
where $f(t)$ is an arbitrary function of the time. 
This space is slightly larger than the space spanned by the conformal Killing vectors 
in which $f(t)$ must be a constant, which is given in Sect.4. 
The extra mode is the lowest mode, $e_{\half (M y)}$, in the mode expansion of $h^{0i}$:
\bb
   h^{0i}=\sum_{M,y} e_{\half (M y)}(t) Y^i_{\half (M y)} 
          +\sum_{J \geq 1} \sum_{M,y} e_{J(My)}(t) Y^i_{J(My)} + \hbox{h.c.}.
\ee     
{}From Eq.(\ref{eigen-v}), we obtain $(\hnabla^2+2)Y^i_{\half (My)}=0$. 
Thus, substituting this mode expansion into the action, we find that   
there is no kinetic term for the mode $e_{\half (M y)}$. Therefore we remove this 
mode by using the residual gauge symmetry. We thus have
\bb
      e_{\half (M y)}=0.
\ee
We call the radiation gauge with this condition the ${\it radiation}^+$ gauge. 
The residual symmetry is now equivalent to the space spanned by the conformal Killing 
vectors.    

\subsection{Mode expansions and canonical commutation relations} 
\noindent

  {}From the gauge fixed action, we can easily obtain the equations of motion. 
The scalar and vector fields are expanded in $X \sim \e^{-i\om t}Y_{JM}$ and 
$A^i \sim \e^{-i\om t}Y^i_{J(My)}$. We then obtain the equations of motion
\bba
   && \left\{ \om^2-(2J+1)^2 \right\} X=0, 
                 \\ 
   && \left\{ \om^2-(2J+1)^2 \right\} A^i=0. 
\eea
Therefore, we obtain $\om=\pm(2J+1)$ for both the scalar 
and gauge fields, and thus they are expanded as  
\bba
   X &=& \sum_{J \geq 0}\sum_M \fr{1}{\sq{2(2J+1)}} \left\{ 
            \vphi_{JM}\e^{-i(2J+1)t}Y_{JM}
            + \vphi^{\dag}_{JM}\e^{i(2J+1)t}Y^*_{JM} \right\}, 
                \label{mode-X}   \\ 
   A^i &=& \sum_{J \geq \half}\sum_{M,y} \fr{1}{\sq{2(2J+1)}} \left\{ 
            q_{J(My)}\e^{-i(2J+1)t}Y^i_{J(My)}
            + q^{\dag}_{J(My)}\e^{i(2J+1)t}Y^{i*}_{J(My)} \right\}.   
                \nonumber  \\  
       &&    \label{mode-A}        
\eea
The fields are normalized such that the canonical commutation relations become
\bba
   \left[ \vphi_{J_1 M_1}, \vphi^{\dag}_{J_2 M_2} \right] &=& \dl_{J_1 J_2}\dl_{M_1 M_2}, 
                       \\ 
   \left[ q_{J_1 (M_1 y_1)}, q^{\dag}_{J_2 (M_2 y_2)} \right] 
        &=& \dl_{J_1 J_2}\dl_{M_1 M_2}\dl_{y_1 y_2},
\eea  
where $\dl_{M_1 M_2}=\dl_{m_1 m_2}\dl_{\prm_1 \prm_2}$.

  For the conformal mode, $\phi \sim \e^{-i\om t}Y_{JM}$, we obtain 
\bb
        \left\{ \om^2-(2J)^2 \right\} \left\{  \om^2-(2J+2)^2 \right\} \phi= 0. 
\ee
Thus, $\om=\pm2J, \pm(2J+2)$, and the conformal mode is expanded as  
\bba
 && \phi = \fr{\pi}{2\sq{b_1}} \biggl[ 2(\hat{q} +\hat{p}t)Y_{00}  
                 \nonumber  \\
   && \qquad 
      + \sum_{J \geq \half}\sum_M \fr{1}{\sq{J(2J+1)}} \left\{ 
            a_{JM}\e^{-i2Jt}Y_{JM}
            + a^{\dag}_{JM}\e^{i2Jt}Y^*_{JM} \right\} 
                     \\ 
   && \qquad  
     + \sum_{J \geq 0}\sum_M \fr{1}{\sq{(J+1)(2J+1)}} \left\{ 
            b_{JM}\e^{-i(2J+2)t}Y_{JM}
            + b^{\dag}_{JM}\e^{i(2J+2)t}Y^*_{JM} \right\} \biggr] ,  
                \nonumber  
\eea
where $Y_{00}=\fr{1}{\sq{{\rm Vol}(S^3)}}=\fr{1}{\sq{2}\pi}$. 
Quantization can be carried out using the standard method of higher-derivative theories. 
We introduce the new variable $\chi=\pd_t \phi$ and rewrite the action in terms of 
the variables $\phi$, $\chi$, $\pd_t \phi$ and $\pd_t \chi$. 
Then, solving the equation $\chi-\pd_t \phi=0$ as a constraint in the 
Dirac procedure, we obtain the commutation relations,  
\bba
     \left[ \hat{q}, \hat{p} \right] &=& i, 
                \\  
     \left[ a_{J_1 M_1}, a^{\dag}_{J_2 M_2} \right] &=& \dl_{J_1 J_2}\dl_{M_1 M_2}, 
                \\
     \left[ b_{J_1 M_1}, b^{\dag}_{J_2 M_2} \right] &=& - \dl_{J_1 J_2}\dl_{M_1 M_2}. 
\eea
Thus, $a_{JM}$ has a positive metric and $b_{JM}$ has a negative metric.

  For the traceless modes, $h^{ij} \sim \e^{-i\om t}Y^{ij}_{J(Mx)}$ 
and $h^{0i} \sim \e^{-i\om t}Y^i_{J(My)}$, in the $\hbox{radiation}^+$ 
gauge, we obtain 
\bba
  && \left\{ \om^2-(2J)^2 \right\} \left\{  \om^2-(2J+2)^2 \right\} h^{ij} = 0, 
                  \\ 
  && (2J-1)(2J+3) \left\{ \om^2 -(2J+1)^2 \right\} h^{0i} =0. 
\eea 
Therefore, $\om=\pm 2J, \pm(2J+2)$ for $h^{ij}$ 
and $\om=\pm(2J+1)$, $J \neq \half$ for $h^{0i}$, and their mode expansions are 
\bba 
  && h^{ij} =
       \fr{1}{4} \sum_{J \geq 1}\sum_{M,x} \fr{1}{\sq{J(2J+1)}} \left\{ 
            c_{J(Mx)}\e^{-i2Jt}Y^{ij}_{J(Mx)}
            + c^{\dag}_{J(Mx)}\e^{i2Jt}Y^{ij*}_{J(Mx)} \right\} 
               \nonumber      \\ 
   && \qquad\quad 
      + \fr{1}{4} \sum_{J \geq 1}\sum_{M,x} \fr{1}{\sq{(J+1)(2J+1)}} \Bigl\{ 
            d_{J(Mx)}\e^{-i(2J+2)t}Y^{ij}_{J(Mx)}
               \label{hij}  \\ 
   && \qquad\qquad\qquad\qquad\qquad\qquad\qquad\qquad
            + d^{\dag}_{J(Mx)}\e^{i(2J+2)t}Y^{ij*}_{J(Mx)} \Bigr\},  
                 \nonumber  \\ 
  && h^{0i} = 
       \fr{1}{2}\sum_{J \geq 1}\sum_{M,y} \fr{1}{\sq{(2J-1)(2J+1)(2J+3)}} 
           \Bigl\{  e_{J(My)}\e^{-i(2J+1)t}Y^i_{J(My)}
                     \nonumber \\ 
   && \qquad\qquad\qquad\qquad\qquad\qquad\qquad\qquad                     
          + e^{\dag}_{J(My)}\e^{i(2J+1)t}Y^{i*}_{J(My)} \Bigr\}. 
          \label{h0i}                              
\eea
As in the case of the conformal mode, the quantization of $h^{ij}$ can be carried out 
by introducing the new variable $\chi^{ij}=\pd_t h^{ij}$. However, $h^{0i}$ 
is second order in time, and therefore it is not necessary to introduce a new variable 
for it. In this way, we obtain the commutation relations    
\bba
    \left[ c_{J_1 (M_1 x_1)}, c^{\dag}_{J_2 (M_2 x_2)} \right] &=& 
       \dl_{J_1 J_2} \dl_{M_1 M_2}\dl_{x_1 x_2},  
                \\ 
    \left[ d_{J_1 (M_1 x_1)}, d^{\dag}_{J_2 (M_2 x_2)} \right] &=& 
       -\dl_{J_1 J_2}\dl_{M_1 M_2}\dl_{x_1 x_2},  
               \\
    \left[ e_{J_1 (M_1 y_1)}, e^{\dag}_{J_2 (M_2 y_2)} \right] &=& 
      -\dl_{J_1 J_2}\dl_{M_1 M_2}\dl_{y_1 y_2}. 
\eea
Thus, the $c_{J (M x)}$ are positive-metric modes, and the $d_{J (M x)}$ and 
$e_{J (M y)}$ are negative-metric modes.

\section{Conformal Algebra on $R\times S^3$}
\setcounter{equation}{0}
\noindent

  In this section, we construct conformal charges and their algebra in explicit forms. 
We know that, from the equation ({\ref{conf-inv}), 
the conformal variation of the partition function with respect to the 
background metric vanishes. 
Thus, conformal invariance means that 
\bba
  0&=& \hg^{\mu\nu} \fr{\dl}{\dl\hg^{\mu\nu}} Z(\hg) 
             \nonumber \\ 
   &=& \int [d\phi dh dA dX]_{\hg} 
            i \hg^{\mu\nu} \fr{\dl I_{\rm CFT}}{\dl\hg^{\mu\nu}} \e^{iI_{\rm CFT}} 
      + \int \hg^{\mu\nu} \fr{\dl}{\dl\hg^{\mu\nu}} [d\phi dh dA dX]_{\hg} 
               \e^{iI_{\rm CFT}}  
              \nonumber   \\
   &=& -i\half \sq{-\hg} \langle \hat{T}^{\mu}_{~\mu} \rangle,
\eea 
where $\hat{T}_{\mu\nu}=T_{\mu\nu}+t_{\mu\nu}$ and 
$T_{\mu\nu}=-\fr{2}{\sq{-\hg}} \fr{\dl I_{\rm CFT}}{\dl\hg^{\mu\nu}}$. 
Here, $t_{\mu\nu}$ is a four-dimensional version of the Schwarz derivative, 
defined in terms of the functional derivative of the Riegert action~\cite{riegert} 
constructed from $\hg$, which gives a constant shift $b_1$ to the Hamiltonian 
in the $R \times S^3$ background~\cite{amm97}.

  Conformal charges are defined by 
\bb
   Q_{\xi}=\int_{S^3} d\Om_3 \xi^{\mu} \hat{T}_{\mu 0}, 
           \label{Qxi}
\ee
where $d\Om_3=d^3 x \sq{\hgm}$ and the quantities $\xi_{\mu}$ are 
conformal Killing vectors on $R \times S^3$ which satisfy the equation 
\bb
    \hnabla_{\mu}\xi_{\nu}+\hnabla_{\nu}\xi_{\mu}
      -\half \hg_{\mu\nu} \hnabla^{\lam}\xi_{\lam}=0.
             \label{ckv}
\ee 
Using Eq.(\ref{ckv}) and the transverse condition for 
the stress tensor, $\hnabla^{\mu}\hat{T}_{\mu\nu}=0$, we can show that 
the charges are conserved: $\dot{Q}_{\xi}=-\fr{1}{3}\int_{S^3} d\Om_3  
\hnabla_i \xi^i \hat{T}^{\mu}_{~\mu}=0$.

  There are 15 conformal Killing vectors on $R \times S^3$: time translation, 
isometries on $S^3$, and special conformal transformations. 
These are discussed below.
\paragraph{Time translation}
Time translation is represented by the vector $\xi_T^{\mu}=(1,0,0,0)$. 
The corresponding charge is the Hamiltonian 
\bb
         H=\int_{S^3} d\Om_3 \hat{T}_{00} .
\ee
\paragraph{Isometries on $S^3$}
The Killing vectors on $S^3$ are solutions of Eq.(\ref{ckv}), which are represented 
by $\xi_R^{\mu}=(0,\xi_R^i)$, where $\hnabla^i \xi_R^j +\hnabla^j \xi_R^i =0$ 
and $\pd_t \xi_R^i =0$. The components $\xi_R^i$ are given by the vector 
harmonics of $J=\half$, $Y^i_{1/2 (My)}$. 
Owing to the characteristics of $J=\half$, the vector harmonics can be expressed 
in terms of the scalar harmonics as 
\bb
    ( \xi^i_R )_{MN} 
      = i \fr{{\rm Vol}(S^3)}{4} \left\{ Y^*_{\half M} \hnabla^i Y_{\half N} 
              - (\hnabla^i Y^*_{\half M}) Y_{\half N} \right\} . 
         \label{xi-R}
\ee
Here, we use this expression to define the rotation generators on $S^3$,
\bb
     R_{MN} =\int_{S^3} d\Om_3 (\xi^i_R)_{MN} \hat{T}_{i0}.   
\ee
{}From the property of the scalar harmonics (\ref{s-harmonics}), 
we can demonstrate the relations 
\bb
     R_{MN}=-\eps_M \eps_N R_{-N-M}, \qquad  \eps_M=(-1)^{m-\prm}.  
             \label{R=R}
\ee  
Thus, only $6$ of these generators are independent. 
\paragraph{Special conformal transformations}
The remaining Killing vectors are the $4+4=8$ vectors given by $\xi_{SC}^{\mu}=(\xi_{SC}^0, \xi_{SC}^i)$, 
\bb 
   \xi_{SC}^0 = \half \sq{{\rm Vol}(S^3)} \e^{it}Y^*_{\half M} , \qquad 
   \xi_{SC}^i = -\fr{i}{2} \sq{{\rm Vol}(S^3)} \e^{it} \hnabla^i Y^*_{\half M} , 
          \label{SC-vector}
\ee
and their complex conjugates, which represent the special conformal transformations.  
Substituting $\xi_{SC}^{\mu}$ into the definition (\ref{Qxi}), 
we obtain the following expression for the charge of special 
conformal transformation: 
\bb
    Q_M = \sq{{\rm Vol}(S^3)} P^{(+)} \int_{S^3} d\Om_3 
            Y^*_{\half M} \hat{T}_{00}, 
            \label{QM}
\ee
where $P^{(+)}=\half \e^{it}(1+i\pd_t)$. The integral over $S^3$ becomes 
functions of $\e^{\pm it}$, so that $P^{(+)}$ selects the $\e^{-it}$ part 
and causes the charge to be time-independent.

  In the following, we will construct these 15 charges, $H$, $R_{MN}$, $Q_M$ 
and $Q^{\dag}_M$, explicitly for scalar fields, vector fields, the conformal 
mode and the traceless mode and show that they form the closed 
algebra~\cite{amm97} defined by the following: 
\bb
     \left[ Q_M, Q^{\dag}_N \right] =2\dl_{MN} H + 2R_{MN}    
         \label{CA}
\ee
and 
\bba
    \left[ H, Q_M \right] &=& -Q_M,  
             \\ 
    \left[ H, R_{MN} \right] &=& \left[ Q_M, Q_N \right] = 0, 
             \\ 
    \left[ Q_M, R_{M_1 M_2} \right] &=& \dl_{M M_2}Q_{M_1} 
                 -\eps_{M_1}\eps_{M_2}\dl_{M -M_1}Q_{-M_2} , 
                 \\ 
    \left[ R_{M_1 M_2}, R_{M_3 M_4} \right]
        &=& \dl_{M_1 M_4} R_{M_3 M_2} -\eps_{M_1}\eps_{M_2} \dl_{-M_2 M_4} R_{M_3 -M_1} 
              \nonumber \\ 
    && - \dl_{M_2 M_3} R_{M_1 M_4} +\eps_{M_1}\eps_{M_2} \dl_{-M_1 M_3} R_{-M_2 M_4} .         
            \label{R-algebra}
\eea

  We can rewrite the algebra among the rotation generators (\ref{R-algebra}) into 
the familiar form of the $SU(2)\times SU(2)$ algebra. 
Let us parametrize the ${\bf 4}$ representation of $SU(2)\times SU(2)$ as 
\bb
  \left\{ \left( \half,\half \right), \left(\half,-\half \right), 
          \left(-\half,\half \right), \left(-\half,-\half \right) \right\} 
  =(1,2,3,4) . 
          \label{parametrization}
\ee 
Then we obtain 
\bba
  &&  R_{14}=R_{41}=R_{23}=R_{32}=0, \qquad R_{11}=-R_{44}, \qquad R_{22}=-R_{33},  
                 \nonumber \\
  &&  R_{12}=R_{34}, \qquad R_{13}=R_{24}, \qquad R_{21}=R_{43}, 
        \qquad  R_{31}=R_{42}.
\eea 
Noting that $R^{\dag}_{MN}=R_{NM}$ and identifying 
\bba
   && A_+ =R_{31}, \qquad A_- =R_{13}=R^{\dag}_{31}, 
        \qquad A_3=\half (R_{11}+R_{22}), 
          \nonumber \\
   && B_+ =R_{21}, \qquad B_- =R_{12}=R^{\dag}_{21}, 
        \qquad B_3=\half (R_{11}-R_{22}),  
\eea
we find that the algebra (\ref{R-algebra}) is equivalent to the 
standard $SU(2)\times SU(2)$ algebra, i.e.,
\bba
  &&  \left[ A_+, A_- \right]=2A_3, \qquad 
      \left[ A_3, A_{\pm} \right]=\pm A_{\pm}, 
         \nonumber \\
  &&  \left[ B_+, B_- \right]=2B_3, \qquad 
      \left[ B_3, B_{\pm} \right]=\pm B_{\pm} ,
\eea
and $A_{\pm,3}$ and $B_{\pm,3}$ commute. The charges $A_{\pm,3}$ $(B_{\pm,3})$ 
act on the left (right) $SU(2)$ index of $M=(m,\prm)$. 

\subsection{Scalar field}
\noindent

  Let us first construct conformal charges for scalar fields. 
The energy density of the scalar field on $R\times S^3$ 
is given by
\bb
   T_{00}= : \half \pd_t X \pd_t X  -\half X \pd_t^2 X  
               +\fr{1}{12}  \hnabla^2(X^2) :, 
\ee
where $::$ indicates the normal ordering. 
Substituting this into the definition of the conformal charge (\ref{QM}) 
and using the mode expansion (\ref{mode-X}) and the property (\ref{s-harmonics}), 
we obtain 
\bba
   Q_M &=&   P^{(+)} \sum_{J_1,M_1}\sum_{J_2,M_2} 
          \fr{1}{4}\sq{\fr{{\rm Vol}(S^3)}{(2J_1+1)(2J_2+1)}} 
            \int_{S^3} d\Om_3 Y^*_{\half M} Y_{J_1 M_1} Y_{J_2 M_2} 
                \nonumber  \\           
       && \times 
         \biggl[  \left\{ -(2J_1+1)(2J_2+1)+(2J_2+1)^2 -\half \right\} 
                 \nonumber  \\ 
       && \qquad \times 
           \left( \vphi_{J_1 M_1}\vphi_{J_2 M_2}\e^{-i(2J_1+2J_2+2)t} 
                  +\tvphi^{\dag}_{J_1 M_1}\tvphi^{\dag}_{J_2 M_2} 
                   \e^{i(2J_1+2J_2+2)t} \right)  
                 \nonumber \\
       && \quad 
           + \left\{ (2J_1+1)(2J_2+1)+(2J_2+1)^2 -\half \right\} 
                 \nonumber  \\ 
       && \qquad \times 
           \left( \vphi_{J_1 M_1}\tvphi^{\dag}_{J_2 M_2}\e^{-i(2J_1-2J_2)t} 
                  +\tvphi^{\dag}_{J_1 M_1}\vphi_{J_2 M_2} 
                   \e^{i(2J_1-2J_2)t} \right)  \biggr] , 
\eea
where $\tvphi_{JM}=\eps_M \vphi_{J-M}$ and $\eps_M$ is defined in (\ref{R=R}).  
Using the $SU(2)\times SU(2)$ Clebsch-Gordan coefficient of  
type $\C$, (\ref{C-coeff}), and the properties, (\ref{triangle-c}) 
and (\ref{property-c}), we find that only the functions of $\e^{\pm it}$  
remain. The projection operator, $P^{(+)}$, selects the $\e^{-it}$ part, and 
we finally obtain~\cite{amm97} 
\bb
    Q_M = \sum_{J \geq 0}\sum_{M_1, M_2} \C^{\half M}_{JM_1, J+\half M_2}  
           \sq{(2J+1)(2J+2)}  \tvphi^{\dag}_{JM_1}\vphi_{J+\half M_2} ,
              \label{QM-s}
\ee
where 
\bba
      \C^{\half M}_{JM_1, J+\half M_2}
       &=&\sq{{\rm Vol}(S^3)} 
         \int_{S^3} d\Om_3 Y^*_{\half M} Y_{J M_1} Y_{J+\half M_2} 
                 \nonumber \\
       &=&\sq{(J+1)(2J+1)} C^{\half m}_{J m_1, J+\half m_2} 
                     C^{\half \prm}_{J \prm_1, J+\half \prm_2} .
\eea

  We can show that the charge (\ref{QM-s}) satisfies the conformal 
algebra (\ref{CA}) with the Hamiltonian  
\bb
    H=\sum_{J \geq 0}\sum_M (2J+1)\vphi^{\dag}_{JM}\vphi_{JM}  
\ee
and the rotation generators on $S^3$, in the parametrization (\ref{parametrization}),  
\bba
   R_{11} &=& \sum_{J > 0}\sum_M (m+\prm)  
             \vphi^{\dag}_{JM}\vphi_{JM} , 
                         \\
   R_{22} &=& \sum_{J > 0}\sum_M (m-\prm)  
             \vphi^{\dag}_{JM}\vphi_{JM} ,   
                        \\
   R_{21} &=& \half \sum_{J > 0}\sum_M \sq{(2J+2-2\prm)(2J+2\prm)} 
             \vphi^{\dag}_{JM}\vphi_{J\bM} , 
                 \\ 
   R_{31} &=& \half \sum_{J > 0}\sum_M \sq{(2J+2-2m)(2J+2m)}  
             \vphi^{\dag}_{JM}\vphi_{J\uM} , 
\eea
where $\bM=(m,\prm-1)$ and $\uM=(m-1,\prm)$.

\subsection{Vector field}
\noindent

  The energy density of the gauge field on $R\times S^3$ 
in the radiation gauge is given by
\bb
   T_{00}= :\half \pd_t A^i \pd_t A_i  -\half A^i (\hnabla^2 -2) A_i  
             +\half \hnabla_i (A_j F^{ij}) :. 
\ee
Substituting this into the definition of the charge (\ref{QM}) and using the mode 
expansion (\ref{mode-A}), we obtain the expression 
\bb
    Q_M = -\sum_{J \geq \half}\sum_{M_1,y_1, M_2, y_2} 
           \D^{\half M}_{J(M_1 y_1), J+\half (M_2 y_2)}  
      \sq{(2J+1)(2J+2)}  \tilde{q}^{\dag}_{J(M_1 y_1)} q_{J+\half (M_2 y_2)} ,
           \label{QM-v}
\ee
where $\tq_{J(My)}=\eps_M q_{J(-My)}$. The overall negative sign comes 
from the relation $Y^{*i}_{J(My)}=-\eps_M Y^i_{J(-My)}$.  
The $\D$-coefficient is defined by  
\bba
  &&  \D^{\half M}_{J(M_1 y_1), J+\half (M_2 y_2)}
       = \sq{{\rm Vol}(S^3)}  
         \int_{S^3} d\Om_3 Y^*_{\half M} Y^i_{J (M_1 y_1)} Y_{i~J+\half (M_2 y_2)} 
                 \nonumber \\
  && \quad 
   =  \sq{J(2J+3)}  C^{\half m}_{J+y_1 m_1,~ J+\half+y_2 m_2} 
              C^{\half \prm}_{J-y_1 \prm_1,~ J+\half-y_2 \prm_2}.
                  \label{D1/2-coeff}
\eea
This coefficient is a special case of the general 
one, $\D^{JM}_{J_1 (M_1 y_1), J_2 (M_2 y_2)}$ defined in Appendix B. 
For the case $J=\half$ given above, we find that the coefficient vanishes 
in the case that the signs of $y_1$ and $y_2$ are different.

  The charge (\ref{QM-v}) satisfies the conformal algebra (\ref{CA}) 
with the Hamiltonian   
\bb
    H= \sum_{J \geq \half}\sum_{M,y} (2J+1) q^{\dag}_{J(My)}q_{J(My)} 
\ee
and the rotation generators
\bba
   R_{11} &=& \sum_{J \geq \half}\sum_{M,y} (m+\prm)  
             q^{\dag}_{J(My)}q_{J(My)} , 
                         \\
   R_{22} &=& \sum_{J \geq \half}\sum_{M,y} (m-\prm)  
             q^{\dag}_{J(My)}q_{J(My)} ,   
             \\
   R_{21} &=& \half \sum_{J \geq \half}\sum_M \sq{(2J+1-2\prm)(2J-1+2\prm)} 
             q^{\dag}_{J(M\half)}q_{J(\bM\half)} 
                 \nonumber  \\ 
  &&  +\half \sum_{J \geq \half}\sum_M \sq{(2J+3-2\prm)(2J+1+2\prm)}  
             q^{\dag}_{J(M-\half)}q_{J(\bM-\half)}, 
                           \\ 
   R_{31} &=& \half \sum_{J \geq \half}\sum_M \sq{(2J+3-2m)(2J+1+2m)}  
             q^{\dag}_{J(M\half)}q_{J(\uM\half)} 
                 \nonumber  \\ 
  &&  +\half \sum_{J \geq \half}\sum_M \sq{(2J+1-2m)(2J-1+2m)}  
             q^{\dag}_{J(M-\half)}q_{J(\uM-\half)}, 
\eea
where $\bM=(m,\prm-1)$ and $\uM=(m-1,\prm)$.

\subsection{Conformal mode}
\noindent

  The stress tensor for the conformal mode is complicated. 
On $R \times S^3$, the energy density is given by  
\bba
   && T_{00}= -\fr{2b_1}{(4\pi)^2} : \biggl\{ 
         (\pd_t^2 \phi)^2 -2\pd_t\phi \pd_t^3\phi -\pd_t^2\phi\hnabla^2\phi 
         +2\pd_t\phi \pd_t\hnabla^2\phi +\phi\pd_t^2\hnabla^2\phi 
            \nonumber \\
   && \qquad\qquad 
        -(\hnabla^2\phi)^2 -4 (\pd_t \phi)^2  
        + \hnabla^2 \left( \fr{2}{3} (\pd_t\phi)^2 -\phi\pd_t^2\phi 
        -\fr{1}{3} \phi\hnabla^2 \phi +\fr{1}{6}\hnabla^2(\phi^2) \right) 
             \nonumber  \\
   && \qquad\qquad  
      + \fr{2}{3}\pd_t^2\hnabla^2 \phi -\fr{2}{3}\hnabla^4 \phi +4\hnabla^2 \phi 
          \biggr\} :,
\eea
and $t_{00}=\fr{b_1}{2\pi^2}$. From this, we obtain~\cite{amm97} 
\bba
    Q_M &=& \left( \sq{2b_1}-i\hat{p} \right) a_{\half M} 
                 \nonumber \\
        && +\sum_{J \geq 0}\sum_M \C^{\half M}_{JM_1, J+\half M_2}  
              \Bigl\{ \a(J)  \tilde{a}^{\dag}_{JM_1} a_{J+\half M_2} 
                  \nonumber  \\
        &&\qquad\qquad
             +\b(J) \tilde{b}^{\dag}_{JM_1} b_{J+\half M_2}  
             +\gm(J) \tilde{a}^{\dag}_{J+\half M_2} b_{J M_1} \Bigr\}, 
\eea
where $\ta_{JM}=\eps_M a_{J-M}$, $\tb_{JM}=\eps_M b_{J-M}$ and 
\bba
    \a(J)&=&\sq{2J(2J+2)},  
          \nonumber  \\
    \b(J)&=&-\sq{(2J+1)(2J+3)}, 
          \label{abc}  \\ 
    \gm(J)&=& 1 . 
             \nonumber  
\eea

  We can show that the charge satisfies the conformal algebra (\ref{CA}). 
When we evaluate the commutator $[Q_M, Q^{\dag}_N]$, the crossing 
relation    
\bb
    \sum_{S} \left\{ 
         \eps_{M_1}\C^{\half M}_{J-\half -M_1, J S}
            \eps_{M_2}\C^{\half N}_{J S, J+\half -M_2} 
     - \C^{\half M}_{J+\half M_2, J S} 
           \C^{\half N}_{J S, J-\half M_1} \right\} =0, 
\ee
which is a special case of (\ref{cross-cc}), 
is useful to show that the off-diagonal parts, $a^{\dag}b$ and $b^{\dag}a$, vanish.  
The diagonal parts, $a^{\dag}a$ and $b^{\dag}b$, produce the Hamiltonian $H$ and 
the rotation generators $R_{MN}$. The Hamiltonian is given by
\bb
   H= \half \hat{p}^2 +b_1 
   +\sum_{J \geq 0}\sum_M \left\{ 
     2J a^{\dag}_{JM}a_{JM}-(2J+2)b^{\dag}_{JM}b_{JM} \right\}, 
\ee 
where the constant shift $b_1$ comes from $t_{00}$.  
The explicit forms of the rotation generators are given in Appendix E.

\subsection{Traceless mode}
\noindent

  Let us determine the conformal charges in the traceless-mode sector.
In this case, the Hamiltonian is easily derived from the action as  
\bba
   H &=& \sum_{J \geq 1}\sum_{M,x} \left\{ 2J c^{\dag}_{J(Mx)}c_{J(Mx)} 
                 -(2J+2)d^{\dag}_{J(Mx)}d_{J(Mx)} \right\} 
                \nonumber  \\
    && -\sum_{J \geq 1}\sum_{M,y} (2J+1) e^{\dag}_{J(My)}e_{J(My)}.
\eea
However, the traceless mode is too complicated to derive an expression of the charge $Q_M$ 
directly from the definition. 
Here we determine it indirectly.

  Our strategy is to assume the form of $Q_M$ and then determine it 
by imposing the condition that it satisfies the conformal algebra. 
Due to the properties that $Q_M$ belongs to the ${\bf 4}$ representation 
of $SU(2)\times SU(2)$ and the relation $[H, Q_M]=-Q_M$, the general form 
is given by    
\bba
    Q_M &=& \sum_{J \geq 1}\sum_{M_1,x_1, M_2,x_2} 
            \E^{\half M}_{J(M_1 x_1), J+\half (M_2 x_2)}  
          \Bigl\{ \a(J) \tilde{c}^{\dag}_{J(M_1 x_1)} c_{J+\half (M_2 x_2)} 
                        \nonumber \\
        &&\qquad\qquad\qquad\qquad\qquad\qquad\qquad 
             +\b(J) \tilde{d}^{\dag}_{J(M_1 x_1)} d_{J+\half (M_2 x_2)}
                        \nonumber \\     
        &&\qquad\qquad\qquad\qquad\qquad\qquad\qquad 
             +\gm(J) \tilde{c}^{\dag}_{J+\half (M_2 x_2)} d_{J (M_1 x_1)} \Bigr\} 
                        \nonumber \\ 
        && + \sum_{J \geq 1}\sum_{M_1,x_1, M_2,y_2} 
            \H^{\half M}_{J(M_1 x_1); J (M_2 y_2)} 
             \Bigl\{ A(J) \tilde{c}^{\dag}_{J(M_1 x_1)} e_{J (M_2 y_2)} 
                   \nonumber  \\ 
        &&\qquad\qquad\qquad\qquad\qquad\qquad\qquad 
                + B(J) \tilde{e}^{\dag}_{J(M_2 y_2)} d_{J (M_1 x_1)} \Bigr\} 
                     \nonumber  \\ 
        && + \sum_{J \geq 1}\sum_{M_1,y_1, M_2, y_2} 
           \D^{\half M}_{J(M_1 y_1), J+\half (M_2 y_2)}  
             C(J) \tilde{e}^{\dag}_{J(M_1 y_1)} e_{J+\half (M_2 y_2)},  
\eea 
where $\tc_{J(Mx)}=\eps_M c_{J(-Mx)}$, $\td_{J(Mx)}=\eps_M d_{J(-Mx)}$ 
and $\te_{J(My)}=\eps_M e_{J(-My)}$.   
Here we use the same notation for the coefficients, $\a$, $\b$ and $\gm$, 
as in the conformal-mode sector. For the time being, we take them to be arbitrary functions. 
Below, we show that they are the same as (\ref{abc}).

  The $SU(2) \times SU(2)$ Clebsch-Gordan coefficients of type $\D$   
are given in (\ref{D1/2-coeff}). 
The $\E$ coefficient is given by the following form:  
\bba
    &&  \E^{\half M}_{J(M_1 x_1), J+\half (M_2 x_2)}
       =\sq{{\rm Vol}(S^3)} 
         \int_{S^3} d\Om_3 Y^*_{\half M} Y^{ij}_{J (M_1 x_1)} Y_{ij J+\half (M_2 x_2)} 
                 \nonumber \\
    && \qquad
       = \sq{(2J-1)(J+2)} C^{\half m}_{J+x_1 m_1, J+\half+x_2 m_2} 
              C^{\half \prm}_{J-x_1 \prm_1, J+\half-x_2 \prm_2}.
\eea  
The $\H$ coefficient is given by  
\bba
    &&  \H^{\half M}_{J(M_1 x_1); J(M_2 y_2)}
       = \sq{{\rm Vol}(S^3)}
           \int_{S^3} d\Om_3 Y^*_{\half M} Y^{ij}_{J (M_1 x_1)} 
              \hnabla_i Y_{j J (M_2 y_2)} 
                 \nonumber \\
    && \qquad
        =- \sq{(2J-1)(2J+3)} 
              C^{\half m}_{J+x_1 m_1, J+y_2 m_2} 
              C^{\half \prm}_{J-x_1 \prm_1, J-y_2 \prm_2}.
\eea
These coefficients are special cases of the general 
ones, $\E^{JM}_{J_1 (M_1 x_1), J_2 (M_2 x_2)}$ 
and  $\H^{JM}_{J_1 (M_1 x_1); J_2 (M_2 y_2)}$, defined in Appendix B. 
For smaller values of $J$, as given above,  we can show that 
they vanish in the cases that the signs of $x_1$ and $x_2 (y_2)$ are different,   
and also that $\H^{\half M}_{J_1 (M_1 x_1); J_2 (M_2 y_2)} \propto \dl_{J_1 J_2}$.

  Let us calculate the commutator of $Q_M$ and $Q^{\dag}_N$. 
We obtain the following form: 
\bba
  && [ Q_M, Q^{\dag}_N ] 
            \nonumber \\
  && =\sum_{J}\sum_{M_1, x_1}\sum_{M_2, x_2} 
            \nonumber \\
  &&\quad
      \Bigl\{ X^{M,N}_{(M_1 x_1), (M_2 x_2)} c^{\dag}_{J(M_1 x_1)}c_{J(M_2 x_2)} 
        -Y^{M,N}_{(M_1 x_1), (M_2 x_2)} d^{\dag}_{J(M_1 x_1)}d_{J(M_2 x_2)} 
            \nonumber  \\ 
  && \quad 
       +Z^{M,N}_{(M_1 x_1), (M_2 x_2)} d^{\dag}_{J-\half (M_1 x_1)}c_{J+\half (M_2 x_2)} 
       +Z^{N,M}_{(M_1 x_1), (M_2 x_2)} d_{J-\half (M_1 x_1)}c^{\dag}_{J+\half (M_2 x_2)}  
         \Bigr\} 
            \nonumber  \\ 
  &&  +\sum_{J}\sum_{M_1, x_1}\sum_{M_2, y_2} 
            \nonumber \\
  &&\quad
      \Bigl\{ U^{M,N}_{(M_1 x_1); (M_2 y_2)} c_{J+\half (M_1 x_1)} e^{\dag}_{J(M_2 y_2)} 
        +U^{N,M}_{(M_1 x_1); (M_2 y_2)} c^{\dag}_{J+\half (M_1 x_1)} e_{J(M_2 y_2)}  
            \nonumber \\ 
  && \quad 
        +V^{M,N}_{(M_1 x_1); (M_2 y_2)} d^{\dag}_{J (M_1 x_1)} e_{J+\half (M_2 y_2)} 
        +V^{N,M}_{(M_1 x_1); (M_2 y_2)} d_{J (M_1 x_1)} e^{\dag}_{J+\half (M_2 y_2)}  
          \Bigr\}
            \nonumber \\   
  && -\sum_{J}\sum_{M_1, y_1}\sum_{M_2, y_2}   
        W^{M,N}_{(M_1 y_1), (M_2 y_2)} e^{\dag}_{J (M_1 x_1)} e_{J (M_2 y_2)}, 
              \label{QQ} 
\eea 
where
\bba 
  && X^{M,N}_{(M_1 x_1), (M_2 x_2)} 
       \nonumber  \\
   && = \sum_{T,x} \biggl\{ 
          \a(J)^2 \eps_{M_1}\E^{\half M}_{J (-M_1 x_1), J+\half (T x)}
           \eps_{M_2}\E^{\half N}_{J+\half (T x), J (-M_2 x_2)}  
         \nonumber \\ 
   &&\qquad\quad
       -\a \left(J-\half \right)^2 \E^{\half M}_{J (M_2 x_2), J-\half (T x)}
                        \E^{\half N}_{J-\half (T x), J (M_1 x_1)}  
         \nonumber \\ 
   &&\qquad\quad 
       -\gm \left(J-\half \right)^2 \eps_{M_1}\E^{\half M}_{J (-M_1 x_1), J-\half (T x)}
                        \eps_{M_2}\E^{\half N}_{J-\half (T x), J (-M_2 x_2)}  
           \biggr\}  
         \nonumber \\ 
   &&\quad 
       -\sum_{V,y} A(J)^2 \eps_{M_1}\H^{\half M}_{J (-M_1 x_1); J (Vy)} 
                          \eps_{M_2}\H^{\half N}_{J (-M_2 x_2); J (Vy)},  
                    \\       
  && Y^{M,N}_{(M_1 x_1), (M_2 x_2)}
       \nonumber  \\   
   && = \sum_{T,x} \biggl\{ 
          \b(J)^2 \eps_{M_1}\E^{\half M}_{J (-M_1 x_1), J+\half (T x)}
                  \eps_{M_2}\E^{\half N}_{J+\half (T x), J (-M_2 x_2)}  
         \nonumber \\ 
   &&\qquad\quad
       -\b \left(J-\half \right)^2 \E^{\half M}_{J (M_2 x_2), J-\half (T x)}
                        \E^{\half N}_{J-\half (T x), J (M_1 x_1)}  
         \nonumber \\ 
   &&\qquad\quad 
       +\gm(J)^2 \E^{\half M}_{J (M_2 x_2), J+\half (T x)}
                     \E^{\half N}_{J+\half (T x), J (M_1 x_1)}  
            \biggr\} 
         \nonumber \\ 
   && \quad 
        -\sum_{V,y} B(J)^2 \H^{\half M}_{J (M_2 x_2); J (Vy)} 
                          \H^{\half N}_{J (M_1 x_1); J (Vy)}, 
                    \\   
  && Z^{M,N}_{(M_1 x_1), (M_2 x_2)}
       \nonumber  \\   
   && = \sum_{T, x} \biggl\{ 
        -\b \left(J-\half \right)\gm(J) 
              \eps_{M_1}\E^{\half M}_{J-\half (-M_1 x_1), J (T x)}
               \eps_{M_2}\E^{\half N}_{J (T x), J+\half (-M_2 x_2)} 
          \nonumber \\
    && \qquad\quad
           -\a(J)\gm \left(J-\half \right) 
                \E^{\half M}_{J+\half (M_2 x_2), J (T x)}
                \E^{\half N}_{J (T x), J-\half (M_1 x_1)}  
             \biggr\}, 
                    \\ 
  && U^{M,N}_{(M_1 x_1); (M_2 y_2)} 
       \nonumber  \\   
   && = \sum_{T,x} \biggl\{ 
         -\a(J)A(J) \E^{\half M}_{J+\half (M_1 x_1), J (T x)}
                          \H^{\half N}_{J (T x); J (M_2 y_2)}  
         \nonumber \\ 
   &&\qquad\quad
       -\gm(J)B(J) \eps_{M_2}\H^{\half M}_{J (T x); J (-M_2 y_2)}
                 \eps_{M_1}\E^{\half N}_{J (T x), J+\half (-M_1 x_1)}  
            \biggr\} 
         \nonumber \\  
   &&\quad
        -\sum_{V,y} C(J)A \left(J+\half \right) 
            \eps_{M_2}\D^{\half M}_{J (-M_2 y_2), J+\half (Vy)} 
            \eps_{M_1}\H^{\half N}_{J+\half (-M_1 x_1); J+\half (Vy)},  
                    \\     
  &&V^{M,N}_{(M_1 x_1); (M_2 y_2)}
       \nonumber  \\   
   && = \sum_{T,x} \biggl\{ 
       -\b(J)B \left(J+\half \right) 
             \eps_{M_1} \E^{\half M}_{J (-M_1 x_1), J+\half (T x)}
             \eps_{M_2} \H^{\half N}_{J+\half (T x); J+\half (-M_2 y_2)}  
         \nonumber \\ 
   &&\qquad\quad
         -\gm(J)A \left(J+\half \right) 
            \H^{\half M}_{J+\half (T x); J+\half (M_2 y_2)}
            \E^{\half N}_{J+\half (T x), J (M_1 x_1)}  
             \biggr\} 
         \nonumber \\  
   &&\quad
       +\sum_{V,y} C(J)B(J) \D^{\half M}_{J+\half (M_2 y_2), J (Vy)} 
                           \H^{\half N}_{J (M_1 x_1); J (Vy)},  
                    \\   
  &&W^{M,N}_{(M_1 y_1), (M_2 y_2)}
       \nonumber  \\   
   && = \sum_{V, y} \biggl\{ C(J)^2 \eps_{M_1}\D^{\half M}_{J (-M_1 y_1), J+\half (V y)}
                          \eps_{M_2}\D^{\half N}_{J+\half (V y), J (-M_2 y_2)} 
          \nonumber \\
    &&\qquad\quad
         -C \left(J-\half \right)^2 
                \D^{\half M}_{J (M_2 y_2), J-\half (V y)}
                \D^{\half N}_{J-\half (V y), J (M_1 y_1)}  \biggr\} 
          \nonumber \\ 
    &&\quad
       +\sum_{T, x} \biggl\{ 
           B(J)^2 \eps_{M_1}\H^{\half M}_{J (T x); J (-M_1 y_1)}
                   \eps_{M_2}\H^{\half N}_{J (T x); J (-M_2 y_2)} 
          \nonumber \\ 
    &&\qquad\quad
              +A(J)^2 \H^{\half M}_{J (T x); J (M_2 y_2)}
                        \H^{\half N}_{J (T x); J (M_1 y_1)}  
              \biggr\}.        
\eea

  In order for the r.h.s. of the commutator (\ref{QQ}) to form the Hamiltonian and 
the rotation generators, the off-diagonal parts, $Z$, $U$ and $V$, must vanish. 
To make $Z$ vanish, the crossing relation  
\bba
   && \sum_{T, x} \Bigl\{ 
         \eps_{M_1}\E^{\half M}_{J-\half (-M_1 x_1), J (T x)}
                          \eps_{M_2}\E^{\half N}_{J (T x), J+\half (-M_2 x_2)} 
          \nonumber \\
    && \qquad
         - \E^{\half M}_{J+\half (M_2 x_2), J (T x)}
               \E^{\half N}_{J (T x), J-\half (M_1 x_1)} \Bigr\} =0  
\eea
is useful. This is a special case of (\ref{cross-ee}).  
We can make $Z$ vanish if the following relation is satisfied:  
\bb
     \a(J)\gm \left( J-\half \right) =- \b \left( J-\half \right) \gm (J). 
                \label{Z1}
\ee 
To make $U$ vanish, we use the crossing relation 
\bba
  && \sum_{T,x} \Bigl\{ 
          \E^{\half M}_{J+\half (M_1 x_1), J (T x)}
                          \H^{\half N}_{J (T x); J (M_2 y_2)}  
         \nonumber \\ 
   &&\qquad
       +\fr{1}{2J} \eps_{M_2}\H^{\half M}_{J (T x); J (-M_2 y_2)}
                      \eps_{M_1}\E^{\half N}_{J (T x), J+\half (-M_1 x_1)}  
            \Bigr\} 
         \nonumber \\  
   && + \fr{(2J-1)(2J+1)}{(2J)^2} \sum_{V,y}  
         \eps_{M_2}\D^{\half M}_{J (-M_2 y_2), J+\half (Vy)} 
            \eps_{M_1}\H^{\half N}_{J+\half (-M_1 x_1); J+\half (Vy)} 
           \nonumber  \\
   &&  =0.  
\eea
This equation is obtained by removing the $\H\cdot \D$ terms 
from Eqs. (\ref{cross-eh}) and (\ref{cross-hd}). 
{}From this, we obtain the relations
\bba
    \fr{1}{2J}\a(J) A(J) &=& \gm(J) B(J), 
        \label{U1} \\ 
    \fr{(2J-1)(2J+1)}{(2J)^2}\a(J) A(J) &=& C(J) A \left( J+\half \right). 
        \label{U2}
\eea
For $V=0$, we use a slightly different crossing relation, 
\bba
   && \sum_{T,x} \Bigl\{ 
         \eps_{M_1} \E^{\half M}_{J (-M_1 x_1), J+\half (T x)}
             \eps_{M_2} \H^{\half N}_{J+\half (T x); J+\half (-M_2 y_2)}  
         \nonumber \\ 
   &&\qquad
       -\fr{1}{2J+3} \H^{\half M}_{J+\half (T x); J+\half (M_2 y_2)}
                      \E^{\half N}_{J+\half (T x), J (M_1 x_1)}  
             \Bigr\} 
         \nonumber \\  
   && + \fr{(2J+2)(2J+4)}{(2J+3)^2} \sum_{V,y} 
          \D^{\half M}_{J+\half (M_2 y_2), J (Vy)}
           \H^{\half N}_{J (M_1 x_1); J (Vy)}  =0.
\eea
{}From this, we obtain the relations
\bba
    \fr{1}{2J+3}\b(J) B \left( J+\half \right) &=& -\gm(J) A \left( J+\half \right), 
           \label{V1}  \\ 
    \fr{(2J+2)(2J+4)}{(2J+3)^2}\b(J) B \left( J+\half \right) &=& -C(J) B(J).  
           \label{V2}
\eea

  The solution of the equations (\ref{Z1}), (\ref{U1}), (\ref{U2}), (\ref{V1}) 
and (\ref{V2}) is not unique, as 
the signs of $\gm$, $A$ and $B$ are indefinite.   
Although these signs are intrinsically fixed when we determine the 
mode expansions (\ref{hij}) and (\ref{h0i}), such information is lost 
when we assume the form of $Q_M$, because they change sign when  
the signs of the modes are changed, i.e., $c_{J(Mx)} \arr -c_{J(Mx)}$, and so on. 
We here fix the convention by choosing $\gm >0$. Then, $A$ and $B$ have the same sign, 
and we take $A, B >0$. 
Also, the signs of $\a$, $\b$ and $C$ are related to 
the metric of the modes $c_{J(Mx)}$, $d_{J(Mx)}$ and $e_{J(My)}$. 
Thus, $\a$ and $\b$ must be positive and  negative, respectively.  
Because of the negative metric of $e_{J(My)}$ and the sign factor of the 
relation $Y^{*i}_{j(My)}=-\eps_M Y^i_{J(-My)}$, 
$C$ is positive.

  {}From Eqs.(\ref{Z1}), (\ref{U1}) and (\ref{V1}), 
we obtain, in the convention mentioned above, 
\bba
     \a(J) &=& \sq{2J(2J+2)} \gm(J) ,
             \nonumber  \\ 
     \b(J) &=& -\sq{(2J+1)(2J+3)} \gm(J), 
             \label{AB}   \\ 
     A(J) &=& \sq{ \fr{2J}{2J+2}} B(J). 
          \nonumber
\eea
{}From Eqs.(\ref{U2}), (\ref{V2}) and (\ref{AB}), we obtain the recursion 
relation
\bb
   A\left( J+\half \right)^2 = \fr{(2J-1)(2J+1)(2J+3)}{(2J)^2(2J+4)} A(J)^2 .
\ee 
By solving this equation and substituting the result into Eqs.(\ref{AB}) 
and (\ref{U2}), we obtain 
\bba
   A(J)&=&\sq{\fr{2J}{(2J-1)(2J+3)}} \lam, 
             \nonumber      \\ 
   B(J)&=&\sq{\fr{2J+2}{(2J-1)(2J+3)}} \lam,
                 \label{ABC}  \\ 
   C(J)&=& \sq{\fr{(2J-1)(2J+1)(2J+2)(2J+4)}{2J(2J+3)}} \gm(J),  
             \nonumber
\eea 
where $\lam$ is a positive constant. The quantities $\lam$ and $\gm(J)$ 
are fixed by the condition that the diagonal 
parts, $X$, $Y$ and $W$, produce $H$ and $R_{MN}$ in the normalization (\ref{CA}) 
when substituting the values (\ref{AB}) and (\ref{ABC}) into them. 
Then, we finally obtain
\bb
     \gm(J)=1, \qquad \lam=\sq{2}.
\ee 
Here, note that  $C(J)$ is different from the factor $\sq{(2J+1)(2J+2)}$  
for vector fields. Specifically, $C(J)$ vanishes at $J=\half$. This is a consequence of 
the ${\rm radiation}^+$ gauge.

\section{Physical States in a Non-critical 3-brane}  
\setcounter{equation}{0}
\noindent

  In this section, we discuss physical states in a non-critical 3-brane. 
We first introduce a conformally invariant vacuum for which all charges vanish. 
This is uniquely given by  
$|\Om \rangle = \e^{-\sq{2b_1}\hat{q}}|0\rangle = \e^{-2b_1\phi(t=i\infty)}|0\rangle$, 
where $|0\rangle$ is the standard Fock vacuum that vanishes when annihilation 
operators act. The correction term denotes the background 
charge.\footnote{
In the Euclidean space,  we can read the background charge as 
$\exp(-\fr{b_1}{(4\pi)^2}\int\sq{\hg}\hat{G}_4 \phi_0)=\exp(-2b_1\chi\phi_0)$ from 
the Wess-Zumino action, $S$, where $\phi_0$ is the zero mode of the conformal mode 
and $\chi=2$ is the Euler number. Now, the background charges are assigned at 
the in- and out-conformally invariant vacuums. 
} 
The physical states are spanned by the Fock space generated on the conformally 
invariant vacuum satisfying the 
conditions~\footnote{
The ``weak" conditions employed in Ref.\cite{amm97} are too strong 
to define physical states. 
} 
\bb
     Q_M |{\rm phys} \rangle = 0
             \label{Q-condition}
\ee
and 
\bb   
     H |{\rm phys} \rangle =4 |{\rm phys} \rangle , \qquad 
     R_{MN} |{\rm phys} \rangle =0,    
             \label{H-R-conditions}
\ee
where $4$ denotes the number of dimensions of the 
world-volume.\footnote{
This originates in the ghost sector concerning the residual gauge symmetry. 
In contrast to the $bc$ ghosts in a non-critical string, it is a quantum mechanical 
system with a finite number of degrees of freedom. 
} 
As in the Gupta-Bleuler procedure, we do not impose a condition concerning $Q^{\dag}_M$. 
Thus, in order to construct physical states, we must find creation operators 
that commute with the charges $Q_M$.

  Physical states can be decomposed into four sectors: scalar fields, vector fields, 
the traceless mode and the conformal mode. 
Each sector is an eigenstate of the Hamiltonian that satisfies the 
condition (\ref{Q-condition}). The conditions (\ref{H-R-conditions}) are imposed last, 
after combining all sectors by adjusting the zero-mode value of the conformal mode. 
In this paper, we briefly discuss the scalar field sector and the traceless-mode sector. 
Detailed arguments on the classification of physical states for the other sectors 
as well as these sectors are given in Ref.\cite{hh}.

  We seek creation operators that commute with $Q_M$. 
Such operators provide building blocks of physical states.  
Let us first calculate the commutators between $Q_M$ and the creation modes.  
Here, we first consider the scalar field sector. 
The commutator of $Q_M$ and $\tvphi^{\dag}_{JM}$ is given by 
\bb
   [ Q_M, \tvphi^{\dag}_{JM_1} ] 
   = \sq{2J(2J+1)}\sum_{M_2} \eps_{M_1} 
     \C^{\half M}_{J -M_1, J-\half M_2}\tvphi^{\dag}_{J-\half M_2}. 
\ee 
Thus, only $\vphi^{\dag}_{00}$ commutes with $Q_M$.

  Consider the operator with the level $H=2J+2$,
\bb
     \tilde{\Phi}^{\dag}_{J_3 M_3} (J) 
     = \sum_{K=0}^J \sum_{M_1,M_2} \fr{f(J,K)}{\sq{(2J-2K+1)(2K+1)}} 
        \C^{J_3 M_3}_{J-K M_1, K M_2} 
        \tvphi^{\dag}_{J-K M_1} \tvphi^{\dag}_{K M_2} ,  
\ee
where $\tilde{\Phi}_{J M} =\eps_M \Phi_{J -M}$. 
The commutator of $Q_M$ and $\tilde{\Phi}^{\dag}_{J_3 M_3}$ is calculated as 
\bba
  && [ Q_M , \tilde{\Phi}^{\dag}_{J_3 M_3}(J) ] 
     = \sum_{K=0}^J \sum_{M_1,M_2} 
      \tvphi^{\dag}_{J-K-\half M_1}\tvphi^{\dag}_{K M_2} 
               \nonumber \\ 
  && \quad \times 
      \sum_S  \biggl\{ f(J,K) \sq{\fr{2J-2K}{2K+1}}
                 \eps_S \C^{\half M}_{J-K-\half M_1, J-K -S} 
                            \C^{J_3 M_3}_{J-K S, K M_2} 
                     \\ 
  && \qquad
         + f \left( J, K+\half \right) \sq{\fr{2K+1}{2J-2K}}
               \eps_S \C^{\half M}_{K M_2, K+\half -S} 
          \C^{J_3 M_3}_{K+\half S, J-K-\half M_1}  \biggr\} . 
                \nonumber 
\eea
Using the crossing relation (\ref{cross-ss}), we can make the r.h.s. vanish 
if and only if $J_3=J$ is an integer 
and $f(J,K)$ satisfies the recursion relation 
\bb
    f \left( J,K+\half \right) = -\fr{2J-2K}{2K+1} f(J,K). 
\ee
Solving this recursion relation, we obtain 
\bb
     f(J,K)=(-1)^{2K} \left( \begin{array}{c}
                                     2J \\
                                     2K 
                                     \end{array} \right) 
                \label{f(J,K)},
\ee 
up to the $J$-dependent normalization. 
Hereafter, we express $\Phi^{\dag}_{J M}(J)$ with (\ref{f(J,K)}) 
as $\Phi^{\dag}_{J M}$. 
If we impose $Z_2$ symmetry, $X \leftrightarrow -X$, 
the operator $(\vphi^{\dag}_{00})^n$ with  odd $n$ is excluded, while that with 
even $n$ is generated from $\Phi^{\dag}_{00}=(\vphi^{\dag}_{00})^2$.

  The operator commuting with the rotation generators, $R_{MN}$, is obtained by 
contracting all indices of multiplicity $M=(m,\prm)$, 
using the $SU(2)\times SU(2)$ Clebsch-Gordan coefficients of type $\C$.    
Thus, the scalar field sector of physical states is, for example, constructed as 
\bb
     \sum_M \tilde{\Phi}^{\dag}_{J M} \Phi^{\dag}_{J M} |\Om \rangle , \qquad  
     \sum_{M_1,M_2,M_3} \C^{J_3 M_3}_{J_1 M_1, J_2 M_2}  
     \Phi^{\dag}_{J_3 M_3} \tilde{\Phi}^{\dag}_{J_1 M_1} 
       \tilde{\Phi}^{\dag}_{J_2 M_2}  |\Om \rangle ,
\ee 
and so on. In this way, we can construct an infinite number of states. 
Because of the factorization property of the $SU(2)\times SU(2)$ Clebsch-Gordan 
coefficients, a general state is factorized into a combination 
of the operators $\Phi^{\dag}_{J M}$. 
Thus, these $\Phi^{\dag}_{J M}$ provide building blocks of the scalar field sector.

  Next, consider the traceless mode sector. 
The commutators between $Q_M$ and the creation modes are given by        
\bba
   \left[ Q_M, \tc^{\dag}_{J(M_1x_1)} \right] 
         &=& \a\left(J-\half\right) \sum_{M_2,x_2} \eps_{M_1} 
           \E^{\half M}_{J(-M_1x_1),J-\half (M_2x_2)} 
           \tc^{\dag}_{J-\half (M_2x_2)} , 
               \nonumber \\ 
   \left[ Q_M, \td^{\dag}_{J(M_1x_1)} \right]
          &=& -\gm (J) \sum_{M_2,x_2} \eps_{M_1} 
           \E^{\half M}_{J(-M_1x_1),J+\half (M_2x_2)} 
           \tc^{\dag}_{J+\half (M_2x_2)}
                \nonumber  \\ 
   &&  -\b\left(J-\half\right) \sum_{M_2,x_2} \eps_{M_1} 
           \E^{\half M}_{J(-M_1x_1),J-\half (M_2x_2)} 
           \td^{\dag}_{J-\half (M_2x_2)} 
                \nonumber \\ 
   &&   -B(J)\sum_{M_2,y_2} \eps_{M_1} 
           \H^{\half M}_{J(-M_1x_1);J (M_2y_2)} 
           \te^{\dag}_{J (M_2y_2)} ,
                    \\ 
   \left[ Q_M, \te^{\dag}_{J(M_1y_1)} \right]
          &=& -A(J) \sum_{M_2,x_2} \eps_{M_1} 
           \H^{\half M}_{J(M_2x_2);J (-M_1y_1)} 
           \tc^{\dag}_{J (M_2x_2)}
                \nonumber  \\ 
   &&  -C\left( J-\half \right) \sum_{M_2,y_2} \eps_{M_1} 
           \D^{\half M}_{J(-M_1y_1),J-\half (M_2y_2)} 
           \te^{\dag}_{J-\half (M_2y_2)}. 
             \nonumber 
\eea
This is one of the most important conclusions of this paper. 
It is worth commenting that the only creation mode that commutes with $Q_M$ 
is $c^{\dag}_{1(Mx)}$. No negative-metric modes, $d^{\dag}$ 
and $e^{\dag}$, commute with $Q_M$. All of them are related to the positive-metric 
mode $c^{\dag}$ through the charge $Q_M$. This implies that 
{\it these negative-metric modes are no longer independent physical modes 
under the physical state condition $Q_M = 0$. Physical states are generated by 
particular combinations of the positive-metric and negative-metric creation modes}. 
Thus, we believe that conformal invariance is a mechanism that 
confines ghosts.

  Let us now look for creation operators that commute with $Q_M$.  
There are only two series for such operatos of level $2J$ with the index of the 
scalar harmonic.  We omit the derivation here. The results are as follows:   
\bba
  && \tilde{A}^{\dag}_{J M} = 
    \sum_{K=1}^{J-1} \sum_{M_1,x_1}\sum_{M_2,x_2} 
      \fr{x(J,K)}{\sq{(2J-2K+1)(2K+1)}} 
            \nonumber \\ 
   &&\qquad\qquad \times
      \E^{JM}_{J-K (M_1x_1), K(M_2,x_2)} 
      \tc^{\dag}_{J-K(M_1x_1)} \tc^{\dag}_{K(M_2x_2)} 
\eea
for $J \geq 2$, with integer $J$, and
\bba
  && \tilde{{\cal A}}^{\dag}_{J-1 M} 
   = \sum_{K=1}^{J-1} \sum_{M_1,x_1}\sum_{M_2,x_2} 
      \fr{x(J,K)}{\sq{(2J-2K+1)(2K+1)}} 
            \nonumber \\ 
  &&\qquad\qquad\qquad\qquad \times 
      \E^{J-1M}_{J-K (M_1x_1), K(M_2,x_2)} 
      \tc^{\dag}_{J-K(M_1x_1)} \tc^{\dag}_{K(M_2x_2)} 
               \nonumber \\
  &&\qquad
     + \sum_{K=1}^{J-2} \sum_{M_1,x_1}\sum_{M_2,x_2} 
      \fr{y(J,K)}{\sq{(2J-2K-1)(2K+1)}} 
            \nonumber \\ 
  &&\qquad\qquad\qquad\qquad \times 
      \E^{J-1M}_{J-K-1 (M_1x_1), K(M_2,x_2)} 
      \td^{\dag}_{J-K-1 (M_1x_1)} \tc^{\dag}_{K(M_2x_2)} 
               \nonumber  \\
  &&\qquad
     +\sum_{K=1}^{J-\fr{3}{2}} \sum_{M_1,x_1}\sum_{M_2,y_2} 
      \fr{w(J,K)}{\sq{(2J-2K)(2K+1)}} 
            \nonumber \\ 
  &&\qquad\qquad\qquad\qquad \times 
      \H^{J-1M}_{J-K-\half (M_1x_1); K(M_2,y_2)} 
      \tc^{\dag}_{J-K-\half (M_1x_1)} \te^{\dag}_{K(M_2y_2)}
               \nonumber \\ 
  &&\qquad 
     +\sum_{K=1}^{J-2} \sum_{M_1,y_1}\sum_{M_2,y_2} 
      \fr{v(J,K)}{\sq{(2J-2K-1)(2K+1)}} 
            \nonumber \\ 
  &&\qquad\qquad\qquad\qquad \times 
      \D^{J-1M}_{J-K-1 (M_1y_1), K(M_2,y_2)} 
      \te^{\dag}_{J-K-1(M_1y_1)} \te^{\dag}_{K(M_2y_2)}  
\eea 
for $J \geq 2$, with integer $J$, where 
\bba 
   && x(J,K)= (-1)^{2K}\sq{ \left( \begin{array}{c}
                                     2J \\
                                     2K 
                                     \end{array} \right)  
                             \left(   \begin{array}{c}
                                     2J-2 \\
                                     2K-1 
                                     \end{array} \right) },   
           \nonumber \\ 
   && y(J,K)=-2(2J-2K-1)x(J,K),
            \\
   && w(J,K)=-2\sq{2}\sq{ \fr{(2J-2K-1)(2J-2K)}{2K(2K-1)(2K+3)} }x(J,K), 
           \nonumber \\ 
   && v(J,K)
           \nonumber \\ 
   && =-\sq{ \fr{(2K-1)(2K+1)(2J-2K-3)(2J-2K-1)}{(2K+3)(2J-2K+1)} } 
               x\left( J,K+\half \right).  
            \nonumber
\eea

  There are other operators with the index of the tensor harmonics up to rank 4~\cite{hh}.    
These operators, along with $c^{\dag}_{1(Mx)}$, provide building blocks of 
the traceless mode sector.   
As discussed above, the operator that commutes with the rotation generators $R_{MN}$ 
are obtained by contracting all indices of multiplicity $M=(m,\prm)$ and $x$ and $y$, 
using the $SU(2)\times SU(2)$ Clebsch-Gordan coefficients.    
The results are encouraging, because the creation operator including the negative-metric 
modes, ${\cal A}^{\dag}_{J-1 M}$, has the advantageous feature that the top term with 
the largest multiplicity is given by the positive-metric modes, $c^{\dag}c^{\dag}$. 
This term may give the greatest contribution to the norm, 
and therefore it is expected to be positive.

  Finally, we briefly discuss the conformal mode sector, which must be managed separately, 
because of the existence of the zero mode. Here we consider the state that depends 
only on the conformal mode, $\e^{ip\hat{q}}|\Om \rangle $.  
The Hamiltonian condition in (\ref{H-R-conditions}) gives the 
equation $\half (p+i\sq{2b_1})^2 +b_1 =4$, 
so that $p$ has the purely imaginary value $-i\fr{\a}{\sq{2b_1}}$  
with $\a=2b_1 \left(1-\sq{1-\fr{4}{b_1}} \right)$, where the fact that $b_1 > 4$ is 
used and the solution that $\a$ approaches $4$ in the classical 
limit $b_1 \arr \infty$ is selected. 
This state, expressed by $\e^{\a\phi(t=i\infty)}|\Om \rangle$, is identified 
with the cosmological constant.  
As in the case of non-critical strings, the conformal mode sector is not normalizable. 
This implies that the partition function is given by a {\it grand canonical ensemble}, 
namely dynamical triangulation~\cite{hey}.  
Now, the chemical potential to stabilize the partition function is the cosmological constant.   
Thus, the divergence of the norm in the conformal mode sector is due to the fact that we do not 
consider a suppression factor of the cosmological constant.   
The other three sectors are normalizable, and they  
are regarded as propagation modes on four-dimensional random surfaces. 

\section{Conclusions}  
\setcounter{equation}{0}
\noindent

  In this paper we proposed a world-volume model of a non-critical 3-brane. 
As in the case of a non-critical string, the kinetic term of the conformal mode is induced 
from the measure as the Wess-Zumino action related to the conformal anomaly.  
We investigated the world-volume dynamics of this model in the strong coupling 
phase, called the conformal-mode dominant phase, where the conformal mode 
fluctuates greatly.  We treated the conformal mode non-perturbatively. 
The model possesses exact conformal invariance and can be described by ${\rm CFT}_4$.  
We canonically quantized the model coupled to scalar fields and vector fields 
on the $R\times S^3$ background.  If coupled to ${\cal N}=4$ ${\rm SYM}_4$, 
this model can be regarded as a world-volume model of a D3-brane.

  In order to define the world-volume dynamics, we introduced the Weyl action, which is 
necessary not only to make the model renormalizable but also 
to remove the world-volume singularity, as discussed in Sect.1.  
The existence of no singularity implies that standard point-like excitations 
are forbidden quantum mechanically,  
because they are essentially black holes in the strong coupling phase of gravity.   
We claimed that the Weyl action is necessary to solve the information loss problem, 
even though it introduces negative-metric modes. 
Furthermore, there is an advantage of the Weyl action, namely,  that  
inflation is induced by (dynamical) scales~\cite{starobinsky, hhr, hamada01b}.

  We showed that conformal invariance 
plays an important role in determining physical states 
in the conformal-mode dominant phase. 
We constructed conformal charges and a conformal algebra and determined   
physical state conditions.    
We found that all negative-metric modes are related to positive-metric modes 
through the conformal charges, and therefore negative-metric modes are themselves 
not independent physical modes. Physical states satisfying the conformal 
invariance conditions are given by particular combinations of positive-metric 
and negative-metric modes. Some definite forms of them were constructed.

\begin{center}
{\bf Acknowledgement}
\end{center}
We wish to thank T. Yukawa for discussions on the problem considered here and 
for informing us of his idea regarding the dynamical triangulation approach to 
four-dimensional random surfaces. 

\begin{center}
{\Large {\bf Appendix}}
\end{center}

\appendix 
\section{Spherical Tensor Harmonics on $S^3$}
\setcounter{equation}{0}
\noindent

  In this appendix, we construct practical forms of symmetric transverse 
traceless (${\rm ST}^2$) tensor harmonics on $S^3$ of arbitrary rank.
The general argument is given in Ref.\cite{ro}. Here, we use the peculiar property 
that the isometry group on $S^3$ is $SO(4)=SU(2)\times SU(2)$. 
The ${\rm ST}^2$ tensor harmonics can be classified according to the representations 
of $SU(2) \times SU(2)$.

  The ${\rm ST}^2$ tensor harmonics of rank $n$, denoted 
by $Y^{i_1 \cdots i_n}_{J (M \eps_n)}$, belong to 
the $(J+\eps_n,J-\eps_n)$ representation 
of $SU(2) \times SU(2)$ for each sign of $\eps_n=\pm \fr{n}{2}$, 
which are the eigenfunctions of the Laplacian on $S^3$,
\bb
     \hnabla^2 Y^{i_1 \cdots i_n}_{J (M \eps_n)}
      =\{ -2J(2J+2)+n \} Y^{i_1 \cdots i_n}_{J (M \eps_n)},  
             \label{eigen-eq}
\ee
where $\hnabla^2 = \hnabla^i \hnabla_i$. The quantity $J ~(\geq \fr{n}{2})$ takes 
integer or half-integer values. $M=(m,\prm)$ represents the multiplicity, 
which take the values  
\bba
  m &=&-J-\eps_n,~ -J-\eps_n+1, \cdots, J+\eps_n -1,~ J+\eps_n, 
             \nonumber \\
  \prm &=&-J+\eps_n,~ -J+\eps_n+1, \cdots, J-\eps_n -1,~ J-\eps_n.  
\eea
The multiplicity of the ${\rm ST}^2$ tensor harmonic of rank $n$ is 
given by the product of the left and right $SU(2)$ 
multiplicities, $(2(J+\eps_n)+1)(2(J-\eps_n)+1)$, 
for each sign of $\eps_n=\pm \fr{n}{2}$, and thus it is totally $2(2J+n+1)(2J-n+1)$ 
for $n \geq 1$. For $n=0$, the multiplicity is given by $(2J+1)^2$.

  In order to construct explicit forms of ${\rm ST}^2$ tensor harmonics on $S^3$, 
we have to specify the coordinate system.
Here we introduce two coordinate systems on $R^4$. 
One is the Cartesian coordinate by $x^{\bmu}$, 
with $\bmu=\bar{0},\bar{1},\bar{2},\bar{3}$,  
and the other is the spherical polar coordinate system, denoted 
by $x^{\mu}=(x^0, x^i)$, with $i=1, 2, 3$ 
and $x^0=r=( x^{\bmu}x_{\bmu})^{1/2}$. 
The metrics in these coordinates take the forms
\bb
    ds_{R^4} = \dl_{\bmu\bnu}dx^{\bmu} dx^{\bnu} 
             = g_{\mu\nu} dx^{\mu} dx^{\nu} ,
\ee
where 
\bb
     g_{\mu\nu}= \left( \begin{array}{cc} 
                         1 & 0  \\
                         0 & r^2 \hgm_{ij} 
                         \end{array}
                  \right),  
\ee 
and $\hgm_{ij}$ is the metric on the unit $S^3$. 
The only nonzero Christoffel symbols for the metric $g_{\mu\nu}$ are 
\bb
    \Gm^0_{ij} = -r\hgm_{ij}, \qquad 
    \Gm^i_{0j} =\Gm^i_{j0}= \fr{1}{r}\dl^i_{~j}, \qquad
    \Gm^i_{jk} = \hat{\Gm}^i_{jk},               
\ee
where $\hat{\Gm}^i_{jk}$ is the Christoffel symbol constructed from the $\hgm_{ij}$.

  We here use the Euler angle parametrization $x^i=(\a,\b,\gm)$, with $i=1,2,3$, 
on $S^3$.  In this parametrization, the metric takes the form   
\bb
  \hgm_{ij}= \fr{1}{4 }\left( \begin{array}{ccc}  
                              1 & 0 & \cos \b \\ 
                              0 & 1 & 0  \\
                        \cos \b & 0 & 1 
                              \end{array} 
                     \right),        
\ee
where $\a$, $\b$ and $\gm$ have the ranges $[0,2\pi]$, $[0,\pi]$ and $[0,4\pi]$, 
respectively. The volume element on the unit $S^3$ is  
\bb
      d \Om_3 = d x^3 \sq{\hgm} = \fr{1}{8}\sin \b d\a d\b d\gm , 
\ee
and the volume is ${{\rm Vol}(S^3)}=2\pi^2$. 
The Christoffel symbols $\hat{\Gm}^i_{jk}=\hat{\Gm}^i_{kj}$ are 
\bba
   && \hat{\Gm}^{\b}_{\a\gm}=\fr{1}{2}\sin \b,
          \nonumber  \\ 
   && \hat{\Gm}^{\a}_{\a\b}=\hat{\Gm}^{\gm}_{\gm\b}=\fr{1}{2}\cot \b, 
                   \\ 
   && \hat{\Gm}^{\gm}_{\a\b}=\hat{\Gm}^{\a}_{\b\gm}=-\fr{1}{2\sin \b}.
           \nonumber 
\eea
The relation between the two coordinate systems $x^{\bmu}$ 
and $x^{\mu}=(r,\a, \b,\gm)$ is given by
\bba
        x^{\bar{0}} &=& r \cos \fr{\b}{2} \cos \fr{1}{2}(\a+\gm) ,
             \nonumber \\
        x^{\bar{1}} &=& r \sin \fr{\b}{2} \sin \fr{1}{2}(\a-\gm) ,
             \nonumber \\ 
        x^{\bar{2}} &=& -r \sin \fr{\b}{2} \cos \fr{1}{2}(\a-\gm) ,
             \nonumber \\
        x^{\bar{3}} &=& -r \cos \fr{\b}{2} \sin \fr{1}{2}(\a+\gm) .
\eea

  The ${\rm ST}^2$ tensor harmonics on $S^3$ can be constructed by restricting 
to $S^3$ tensors on $R^4$. 
We first construct scalar harmonics.  
Some important formulae to define ${\rm ST}^2$ tensors of higher rank 
are introduced here. Then, we construct higher rank tensors.

\paragraph{Scalar harmonics}
The scalar harmonics on $S^3$, denoted by $Y_{JM}$, which satisfy 
the equation $\hnabla^2 Y_{JM} =-2J(2J+2) Y_{JM}$, belong to 
the $(J,J)$ representation of the isometry group $SU(2) \times SU(2)$. 
It is well known that the the Wigner $D$ function $D^J_{m \prm}(x^i)$ in the 
Euler angle parametrization is an eigenfunction of the Laplacian on $S^3$. 
Thus, the scalar harmonics $Y_{JM}$ on $S^3$ can be expressed 
in terms of the Wigner $D$ function, where $M=(m,\prm)$ and $m=-J, \cdots, J$, 
$\prm=-J, \cdots, J$.  
If we fix the normalization as 
\bb
    \int_{S^3} d\Om_3 Y^*_{J_1 M_1} Y_{J_2 M_2} = \dl_{J_1 J_2} \dl_{M_1 M_2},      
     \qquad \dl_{M_1 M_2}=\dl_{m_1 m_2}\dl_{\prm_1 \prm_2}, 
            \label{s-norm}
\ee
we obtain  
\bb
  Y_{JM} = \sq{ \fr{2J+1}{{\rm Vol}(S^3)} } D^J_{m \prm}, \qquad 
  Y^*_{JM}=\eps_M Y_{J-M},     
              \label{s-harmonics}
\ee
where $\eps_M=(-1)^{m-\prm}$.

  The Wigner $D$-function/scalar harmonics can be constructed 
if we imbed $S^3$ into the four-dimensional Euclidean space $R^4$. 
Let $\tau_{\bmu_1 \cdots \bmu_n}$ be a symmetric traceless 
rank-$n$ tensor on $R^4$ whose components are constant. 
Then, the Wigner $D$ function can be  
expressed in terms of a polynomial in the coordinates $x^{\bmu}$:  
\bba
    && D^J_{m \prm} = \fr{1}{r^{2J}} x^{\bmu_1} \cdots x^{\bmu_{2J}} 
                       ( \tau_{\bmu_1 \cdots \bmu_{2J}} )_{m \prm}, 
                       \\ 
    && r=(x^{\bmu}x_{\bmu})^{1/2}.
\eea
{}From the second equation in (\ref{s-harmonics}), $\tau_{\bmu_1 \cdots \bmu_n}$ satisfies 
\bb
     ( \tau_{\bmu_1 \cdots \bmu_n} )^*_{m \prm} 
      =\eps_M ( \tau_{\bmu_1 \cdots \bmu_n} )_{-m -\prm}. 
             \label{tau*}
\ee

  The tensor $\tau_{\bmu_1 \cdots \bmu_n}$ is used to define the ${\rm ST}^2$ spherical tensor 
harmonics of rank $n$ below.  Here we give the explicit form of $\tau_{\bmu}$. 
The Wigner $D$ function for $J=\half$ is given by   
\bb
   r D^{\half}_{m \prm} = \left( 
   \begin{array}{cc} 
      r\cos \fr{\b}{2} \e^{-\fr{i}{2}(\a+\gm)} & -r\sin \fr{\b}{2}\e^{-\fr{i}{2}(\a-\gm)} \\ 
      r\sin \fr{\b}{2} \e^{\fr{i}{2}(\a-\gm)} & r\cos \fr{\b}{2}\e^{\fr{i}{2}(\a+\gm)} 
   \end{array} \right) 
   =  \left( 
   \begin{array}{cc} 
      x^{\bar{0}}+ix^{\bar{3}} & x^{\bar{2}}+ix^{\bar{1}} \\ 
      -x^{\bar{2}}+ix^{\bar{1}} & x^{\bar{0}}-ix^{\bar{3}} 
   \end{array} \right),   
\ee
where $m, \prm= \half, -\half$. 
This is identified with $x_{\bmu}(\tau^{\bmu})_{m \prm}$, so that we obtain 
\bb
  \tau^{\bar{0}}= \left( 
      \begin{array}{cc} 
       1 & 0 \\ 
       0 & 1  
      \end{array} \right), \quad 
  \tau^{\bar{1}}= \left( 
      \begin{array}{cc} 
       0 & i \\ 
       i & 0  
      \end{array} \right), \quad 
  \tau^{\bar{2}}= \left( 
      \begin{array}{cc} 
       0 & 1 \\ 
       -1 & 0  
      \end{array} \right), \quad  
  \tau^{\bar{3}}= \left( 
      \begin{array}{cc} 
       i & 0 \\ 
       0 & -i  
      \end{array} \right),  
\ee
where the $(1,1)$ compornents are $(\tau^{\bmu})_{\half \half}$.

  For $J=1$, the Wigner $D$ function is  
\bba
 &&  D^1_{m \prm}=  \left( 
   \begin{array}{ccc} 
   \fr{1+\cos\b}{2} \e^{-i(\a+\gm)} & -\fr{\sin\b}{\sq{2}}\e^{-i\a} 
                            & \fr{1-\cos\b}{2} \e^{-i(\a-\gm)}  \\ 
    \fr{\sin\b}{\sq{2}}\e^{-i\gm} & \cos\b & -\fr{\sin\b}{\sq{2}}\e^{i\gm} \\ 
    \fr{1-\cos\b}{2} \e^{i(\a-\gm)} & \fr{\sin\b}{\sq{2}}\e^{i\a} 
                            & \fr{1+\cos\b}{2} \e^{i(\a+\gm)}
   \end{array} \right), 
\eea
where $m, \prm=1,0,-1$.
{}From this expression, we can easily obtain $(\tau^{\bmu\bnu})_{m \prm}$.

  For later use, we here give some important formulae satisfied by $\tau_{\bmu}$ 
and $\tau_{\bmu\bnu}$. 
{}From the normalization of the scalar harmonics, we obtain the expression 
\bba
  && \dl_{m_1 m_2}\dl_{\prm_1 \prm_2} 
      = \int_{S^3} d\Om_3 Y^*_{J M_1} Y_{J M_2} 
            \\ 
  && = \fr{2J+1}{{\rm Vol}(S^3)} \int_{S^3} d\Om_3  \fr{1}{r^{4J}}
        x^{\bmu_1} \cdots x^{\bmu_{2J}} x^{\bnu_1} \cdots x^{\bnu_{2J}} 
        (\tau_{\bmu_1 \cdots \bmu_{2J}})^*_{m_1 \prm_1} 
        (\tau_{\bnu_1 \cdots \bnu_{2J}})_{m_2 \prm_2}. 
        \nonumber 
\eea
For $J=\half$ and $J=1$, using the integral formulae
\bba
    \int_{S^3} d\Om_3 x^{\bmu}x^{\bnu} 
    &=& \fr{1}{4} \dl^{\bmu\bnu}\int_{S^3} d\Om_3 x^2 
        = \fr{r^2 {\rm Vol}(S^3)}{4} \dl^{\bmu\bnu}, 
                  \\ 
    \int_{S^3} d\Om_3 x^{\bmu}x^{\bnu} x^{\blam}x^{\bs}
       &=& \fr{r^4 {\rm Vol}(S^3)}{24} (\dl^{\bmu\bnu}\dl^{\blam\bs} 
                  +\dl^{\bmu\blam}\dl^{\bnu\bs}
                  +\dl^{\bmu\bs}\dl^{\bnu\blam}),                              
\eea
we obtain the equations
\bba
     (\tau^{\bmu})^*_{m_1 \prm_1} (\tau_{\bmu})_{m_2 \prm_2} 
     = 2 \dl_{M_1 M_2},    
         \label{tau-v} \\
     (\tau^{\bmu\bnu})^*_{m_1 \prm_1} (\tau_{\bmu\bnu})_{m_2 \prm_2} 
     = 4 \dl_{M_1 M_2}.  
         \label{tau-t}   
\eea

  Furthermore, using the $\C$ coefficient calculated in the next section, 
we obtain the equations
\bb
    (\tau_{\bmu})^*_{m_1\prm_1} (\tau^{\bmu\bnu})_{m\prm} 
     (\tau_{\bnu})_{m_2\prm_2} 
     = 2\sq{3} \C^{\half M_1}_{1 M, \half M_2}. 
\ee
As a variant of these equations, 
using $\sum_{m\prm}(\tau^{\bmu})^*_{m\prm}(\tau^{\bnu})_{m\prm} =2\dl^{\bmu\bnu}$, 
we obtain
\bb
   (\tau^{\bmu\bnu})_{m\prm}=\fr{\sq{3}}{2}\sum_{M_1,M_2} 
    \C^{1M}_{\half M_1, \half M_2} 
     (\tau^{\bmu})_{m_1\prm_1}(\tau^{\bnu})_{m_2\prm_2} . 
      \label{tau3}
\ee
In this way, higher rank tensors can be constructed 
from compositions of the tensors $\tau^{\bmu}$.

  Finally,  we give the important equation for the scalar 
harmonic with $J=\half$,    
\bb
  \hnabla^i \hnabla^j Y_{\half M} =\fr{1}{3}\hgm^{ij}\hnabla^2 Y_{\half M} 
  = -\hgm^{ij} Y_{\half M} .  
         \label{special-s}
\ee

\paragraph{Vector harmonics}
Combining the scalar harmonics, the Clebsch-Gordan coefficients 
and $\tau^{\bmu}$, we construct the vector harmonisc on $S^3$, 
in the Cartesian coordinate system on $R^4$, as 
\bb
   Y^{\bmu}_{J (My)} 
   = \fr{1}{\sq{2}} \fr{1}{r}\sum_{N,S} 
      C^{J+y m}_{J n, \half s} C^{J-y \prm}_{J \prn, \half \prs} 
       Y_{J N} (\tau^{\bmu})_{s\prs}, 
\ee
where $N=(n,\prn)$, $S=(s,\prs)$ and $y=\pm \half$. 
Using Eqs.(\ref{s-norm}), (\ref{tau-v}) and (\ref{CC}), 
we can show that these harmonics satisfy the normalization
\bb
   \int_{S^3} d\Om_3 Y^{\bmu*}_{J_1 (M_1y_1)} Y_{\bmu J_2 (M_2y_2)} 
    = \fr{1}{r^2} \dl_{J_1 J_2} \dl_{M_1 M_2}\dl_{y_1 y_2},   
\ee
and, from the relations (\ref{s-harmonics}), (\ref{tau*}) and (\ref{property-C}), 
we have
\bb
        Y^{\bmu*}_{J(My)}=-\eps_M Y^{\bmu}_{J(-My)} . 
\ee
Furthermore, using the 
equation $x^{\bmu}(\tau_{\bmu})_{s \prs}=r\sq{\fr{{\rm Vol}(S^3)}{2}} Y_{\half S}$ 
and the property of the $D$ function given in (\ref{D3}), we obtain 
\bb
    x_{\bmu} Y^{\bmu}_{J (My)}
    =\fr{{\rm Vol}(S^3)}{2} \sum_{N,S} 
       C^{J+y m}_{Jn,\half s} 
       C^{J-y \prm}_{J\prn,\half \prs} Y_{JN}Y_{\half S} =0. 
          \label{transv-v}
\ee
This implies that, when going to spherical polar coordinates on $R^4$, 
the $r=x^0$ component of $Y^{\mu}_{J (My)}$ vanishes. 
Thus, using the relation $Y^{\bmu}Y_{\bmu}=Y^{\mu}Y_{\mu}=\fr{1}{r^2}Y^iY_i$, 
the normalization is equivalent to   
\bb
   \int_{S^3} d\Om_3 Y^{i*}_{J_1 (M_1y_1)} Y_{i J_2 (M_2y_2)} 
    = \dl_{J_1 J_2} \dl_{M_1 M_2}\dl_{y_1 y_2}.  
\ee
We can also show that $Y^i_{J(My)}$ satisfies the transverse 
condition, $\hnabla_i Y^i_{J (My)}=0$. By differentiating Eq.(\ref{transv-v}), 
we obtain the expression
\bb
       Y^{\bmu}_{J(My)} = -x_{\blam}\pd^{\bmu}Y^{\blam}_{J(My)} .
              \label{another-v}
\ee 
{}From this, we find 
that $\pd_{\bmu}Y^{\bmu}=-\half x_{\blam}\Box Y^{\blam} \propto x_{\blam}Y^{\blam}=0$.  
Thus, $\pd_{\bmu}Y^{\bmu}=\nabla_{\mu}Y^{\mu}=\fr{1}{r^2}\hnabla_i Y^i=0$.

  Using the relation between the two coordinate systems,  
\bb
            Y_{\mu J(My)}=\fr{\pd x^{\bmu}}{\pd x^\mu}Y_{\bmu J(My)},  
\ee 
we can obtain explicit forms of the vector harmonics in spherical polar 
coordinates. As mentioned above, the radial component vanishes,  
\bb
    Y_{r J(My)} = 0,  
\ee 
where $y=\pm \half$. The angular components of $y=\half$ are 
\bba
    Y_{\a J(M \half)} &=& 
        \fr{i}{2\sq{2}} \sq{ \fr{(2J+2m+1)(2J-2m+1)}{(2J+1){\rm Vol}(S^3)} } 
          D^{J-\half}_{m \prm} ,
         \nonumber  \\
     Y_{\b J(M \half)} &=& 
        \fr{1}{\sq{2}(2J+1)} \fr{1}{\sin\b} \Biggl\{ 
          m \sq{ \fr{(2J+2\prm+1)(2J-2\prm+1)}{(2J+1){\rm Vol}(S^3)} } 
              D^{J+\half}_{m \prm}  
                \nonumber  \\ 
      && \qquad\qquad\qquad\qquad           
          -\prm \sq{ \fr{(2J+2m+1)(2J-2m+1)}{(2J+1){\rm Vol}(S^3)} } 
              D^{J-\half}_{m \prm} \Biggr\} ,
         \nonumber  \\               
    Y_{\gm J(M \half)} &=& 
        \fr{i}{2\sq{2}} \sq{ \fr{(2J+2\prm+1)(2J-2\prm+1)}{(2J+1){\rm Vol}(S^3)} } 
          D^{J+\half}_{m \prm}                      
\eea
and those of $y=-\half$ are 
\bba
    Y_{\a J(M -\half)} &=& 
        \fr{i}{2\sq{2}} \sq{ \fr{(2J+2m+1)(2J-2m+1)}{(2J+1){\rm Vol}(S^3)} } 
          D^{J+\half}_{m \prm} ,
         \nonumber  \\
     Y_{\b J(M -\half)} &=& 
        \fr{1}{\sq{2}(2J+1)} \fr{1}{\sin\b} \Biggl\{ 
          \prm \sq{ \fr{(2J+2m+1)(2J-2m+1)}{(2J+1){\rm Vol}(S^3)} } 
            D^{J+\half}_{m \prm}  
                \nonumber  \\ 
      && \qquad\qquad\qquad\qquad          
          -m \sq{ \fr{(2J+2\prm+1)(2J-2\prm+1)}{(2J+1){\rm Vol}(S^3)} } 
            D^{J-\half}_{m \prm} \Biggr\} ,
         \nonumber  \\               
    Y_{\gm J(M -\half)} &=& 
        \fr{i}{2\sq{2}} \sq{ \fr{(2J+2\prm+1)(2J-2\prm+1)}{(2J+1){\rm Vol}(S^3)} } 
          D^{J-\half}_{m \prm} .                     
\eea
These expressions are not used in this paper.

  For the special case $J=\half$, the vector harmonics satisfy
\bb
     \hnabla^{(i}Y^{j)}_{\half (My)}
     =\half \left( \hnabla^i Y^j_{\half (My)}+\hnabla^j Y^i_{\half (My)} \right) 
     =0.
            \label{special-v}
\ee 
This equation expresses the fact that the vector harmonics $Y^i_{\half (My)}$ 
are the Killing vectors on $S^3$.

\paragraph{Tensor harmonics} 
As in the case of the vector harmonics, 
the tensor harmonics of rank 2 in Cartesian coordinates on $R^4$ can be 
constructed, using the symmetric traceless tensor $\tau^{\bmu\bnu}$, as 
\bb
   Y^{\bmu\bnu}_{J (Mx)} 
   = \half \fr{1}{r^2} \sum_{N,T} 
      C^{J+x m}_{J n, 1 t} C^{J-x \prm}_{J \prn, 1 \prt} 
        Y_{J N} (\tau^{\bmu\bnu})_{t\prt}, 
             \label{t-harmonics}
\ee
where $N=(n,\prn)$, $T=(t,\prt)$ and $x=\pm 1$. 
Using Eqs.(\ref{s-norm}), (\ref{tau-t}) and (\ref{CC}), we can show that 
this satisfies the normalization
\bb
   \int_{S^3} d\Om_3 Y^{\bmu\bnu*}_{J_1 (M_1x_1)} Y_{\bmu\bnu J_2 (M_2x_2)} 
    = \fr{1}{r^4}\dl_{J_1 J_2} \dl_{M_1 M_2}\dl_{x_1 x_2},   
\ee
and, from the relations (\ref{s-harmonics}), (\ref{tau*}) and (\ref{property-C}), 
we have
\bb
     Y^{\bmu\bnu*}_{J (Mx)} = \eps_M Y^{\bmu\bnu}_{J (-Mx)} .
\ee

  As in the case of the vector harmonics, we can easily show that the 
tensors given by (\ref{t-harmonics}) satisfy the equation   
\bb
    x_{\bmu} x_{\bnu} Y^{\bmu\bnu}_{J (Mx)}
    =\fr{{\rm Vol}(S^3)}{2} \sum_{N,T,U,S} C^{J+x m}_{J n, 1 t} C^{1t}_{\half u, \half s}
        C^{J-x \prm}_{J \prn, 1 \prt} C^{1\prt}_{\half \pru, \half \prs} 
        Y_{JN}Y_{\half U} Y_{\half S}  
    =0, 
\ee
where $U=(u,\pru)$, and the properties of the $D$ function (\ref{D4}) have been used. 
Furthermore, mapping to spherical polar coordinates using the relation 
$Y_{\mu\nu J(Mx)}=\fr{\pd x^{\bmu}}{\pd x^\mu} 
\fr{\pd x^{\bnu}}{\pd x^\nu}  Y_{\bmu\bnu J(Mx)}$, 
we can directly show that the $Y_{r\mu}$ components vanish. This implies that 
the equation  
\bb
    x_{\bmu} Y^{\bmu\bnu}_{J (Mx)}=0 
      \label{transv-t}
\ee
is satisfied. We can also directly see that the harmonics $Y^{ij}_{J(Mx)}$ 
derived in this way satisfy the transverse condition and the 
eigenequation (\ref{eigen-eq}) for $n=2$.

  We can easily generalize the construction of the vector and tensor  
harmonics mentioned above to any rank as 
\bb
   Y^{\bmu_1 \cdots \bmu_n}_{J (M \eps_n)} 
   \propto \sum_{P,U} 
      C^{J+\eps_n m}_{J p, \fr{n}{2} u} 
      C^{J-\eps_n \prm}_{J \prp, \fr{n}{2} \pru} 
       Y_{J P} (\tau^{\bmu_1 \cdots \bmu_n})_{u\pru} ,
\ee
where $P=(p,\prp)$ and $\eps_n=\pm \fr{n}{2}$.

\section{$SU(2) \times SU(2)$ Clebsch-Gordan Coefficients}
\setcounter{equation}{0}
\noindent

  In this section we calculate several $SU(2) \times SU(2)$ Clebsch-Gordan 
coefficients defined by the integrals of three products of ${\rm ST}^2$ tensor harmonics.

\paragraph{$\C$ coefficients}
The most simple coefficient is given by the integral of a product of three scalar harmonics. 
This coefficient can easily be calculated using the following property of 
the Wigner $D$ function: 
\bba
     \C^{JM}_{J_1M_1,J_2M_2} 
      &=& \sq{{\rm Vol}(S^3)}
         \int_{S^3} d\Om_3 Y^*_{JM}Y_{J_1M_1}Y_{J_2M_2} 
              \label{C-coeff} \\ 
      &=& \sq{\fr{(2J_1+1)(2J_2+1)}{2J+1}} C^{Jm}_{J_1m_1,J_2m_2} 
                           C^{J\prm}_{J_1\prm_1,J_2\prm_2} .        
              \nonumber 
\eea
This coefficient vanishes unless the triangular conditions 
\bb
     |J_1 -J_2| \leq J \leq J_1 +J_2 , 
       \label{triangle-c} 
\ee
with integer $J+J_1+J_2$, and the requirement $M=M_1+M_2$ are satisfied. 
{}From the definition (\ref{C-coeff}), we can easily see that the $\C$ coefficients 
satisfy the relations 
\bb
    \C^{JM}_{J_1M_1,J_2M_2}=\C^{JM}_{J_2M_2,J_1M_1} 
    =\C^{J-M}_{J_1-M_1,J_2-M_2}
    = \eps_{M_2}\C^{J_1M_1}_{JM,J_2-M_2}.
           \label{property-c}
\ee

\paragraph{$\D$ coefficients}
Consider the $SU(2)\times SU(2)$ Clebsch-Gordan coefficients including vector harmonics. 
The most simple one is  
\bba
    \D^{JM}_{J_1(M_1y_1),J_2(M_2y_2)} 
      &=& \sq{{\rm Vol}(S^3)}
         \int_{S^3} d\Om_3 Y^*_{JM}Y^i_{J_1(M_1y_1)}Y_{i J_2(M_2y_2)}  
             \label{D-coeff}  \\    
     &=& r^2 \sq{{\rm Vol}(S^3)} 
         \int_{S^3} d\Om_3 Y^*_{JM}Y^{\bmu}_{J_1(M_1y_1)}Y_{\bmu J_2(M_2y_2)}.
          \nonumber 
\eea
Using the result for three scalar harmonics, (\ref{C-coeff}), and 
the formula (\ref{CCC}) relating three Clebsch-Gordan coefficients  for 
the left and right $SU(2)$ parts separately, we obtain
\bba
  && \D^{JM}_{J_1(M_1y_1),J_2(M_2y_2)} 
   = -\sq{ \fr{2J_1(2J_1+1)(2J_1+2)2J_2(2J_2+1)(2J_2+2)}{2J+1} } 
          \nonumber \\
  && \qquad\qquad\qquad\quad \times 
    \left\{ \begin{array}{ccc}
              J   & J_1     & J_2 \\
            \half & J_2+y_2 & J_1+y_1 
            \end{array} \right\}      
    \left\{ \begin{array}{ccc}
              J   & J_1     & J_2 \\
            \half & J_2-y_2 & J_1-y_1 
            \end{array} \right\}            
               \nonumber \\ 
  && \qquad\qquad\qquad\quad \times 
      C^{Jm}_{J_1+y_1 m_1,J_2+y_2 m_2} 
      C^{J\prm}_{J_1-y_1 \prm_1,J_2-y_2 \prm_2}.
\eea
The triangular condition for the standard Clebsch-Gordan coefficients implies that  
this coefficient vanishes unless the triangular conditions  
\bba
     |J_1 -J_2| \leq &J& \leq J_1 +J_2 -1 \qquad \hbox{for $y_1=y_2$}, 
               \nonumber  \\
     |J_1 -J_2|+1 \leq &J& \leq J_1 +J_2  \qquad \hbox{for $y_1 \neq y_2$},  
\eea
with integer $J+J_1+J_2$, and the requirement $M=M_1+M_2$ are satisfied. 
Also, the $\D$ coefficients satisfy the relations
\bb
    \D^{JM}_{J_1(M_1y_1),J_2(M_2y_2)} =\D^{JM}_{J_2(M_2y_2),J_1(M_1y_1)}
    =\D^{J-M}_{J_1(-M_1y_1),J_2(-M_2y_2)} .
\ee

\paragraph{$\E$ coefficients}
The most simple coefficient including tensor harmonics is 
\bba
    \E^{JM}_{J_1(M_1x_1),J_2(M_2x_2)} 
     &=&  \sq{{\rm Vol}(S^3)} 
         \int_{S^3} d\Om_3 Y^*_{JM}Y^{ij}_{J_1(M_1x_1)}Y_{ij J_2(M_2x_2)}  
            \label{E-coeff}  \\ 
     &=& r^4 \sq{{\rm Vol}(S^3)} 
    \int_{S^3} d\Om_3 Y^*_{JM}Y^{\bmu\bnu}_{J_1(M_1x_1)}Y_{\bmu\bnu J_2(M_2x_2)}.  
          \nonumber 
\eea
As in the calculation of the $\D$ coefficients, 
using Eqs. (\ref{C-coeff}) and (\ref{CCC}), we obtain
\bba
  && \E^{JM}_{J_1(M_1x_1),J_2(M_2x_2)} 
              \nonumber \\ 
  &&  = \sq{ \fr{(2J_1-1)(2J_1+1)(2J_1+3)(2J_2-1)(2J_2+1)(2J_2+3)}{2J+1} } 
             \nonumber \\
  && \quad \times  
    \left\{ \begin{array}{ccc}
              J   & J_1     & J_2 \\
              1 & J_2+x_2 & J_1+x_1 
            \end{array} \right\}      
     \left\{ \begin{array}{ccc}
              J   & J_1     & J_2 \\
              1 & J_2-x_2 & J_1-x_1 
            \end{array} \right\}            
               \nonumber \\ 
  && \quad \times 
      C^{Jm}_{J_1+x_1 m_1,J_2+x_2 m_2} 
      C^{J\prm}_{J_1-x_1 \prm_1,J_2-x_2 \prm_2}.
\eea
This coefficient vanishes unless the triangular conditions
\bba
     |J_1 -J_2| \leq &J& \leq J_1 +J_2 -2 \qquad \hbox{for $x_1=x_2$}, 
               \nonumber  \\
     |J_1 -J_2|+2 \leq &J& \leq J_1 +J_2  \qquad \hbox{for $x_1 \neq x_2$}, 
\eea
with integer $J+J_1+J_2$, and the requirement $M=M_1+M_2$ are satisfied.
Also, the relations
\bb
    \E^{JM}_{J_1(M_1x_1),J_2(M_2x_2)}=\E^{JM}_{J_2(M_2x_2),J_1(M_1x_1)}
    =\E^{J-M}_{J_1(-M_1x_1),J_2(-M_2x_2)} 
\ee
are satisfied.

\paragraph{$\G$ coefficients} 
Consider the $SU(2) \times SU(2)$ Clebsch-Gordan coefficients with a derivative. 
The most simple one is constructed from scalar and vector harmonics as  
\bba
    \G^{JM}_{J_1(M_1y_1);J_2 M_2} 
      &=& \sq{{\rm Vol}(S^3)} 
         \int_{S^3} d\Om_3 Y^*_{JM}Y^i_{J_1(M_1y_1)}\hnabla_i Y_{J_2 M_2} 
             \label{G-coeff} \\ 
      &=& -r^2 \sq{{\rm Vol}(S^3)} 
         \int_{S^3} d\Om_3 (\pd_{\bmu}Y^*_{JM}) Y^{\bmu}_{J_1(M_1y_1)}Y_{J_2 M_2}.
             \nonumber  
\eea
This coefficient is evaluated as follows. 
Recall that vector harmonics satisfy the 
condition $x_{\blam}Y^{\blam}_{J(My)}=0$, i.e. (\ref{transv-v}). 
By differentiating this condition, we obtain the expression
\bba
       Y^{\bmu}_{J(My)} &=& -x_{\blam}\pd^{\bmu}Y^{\blam}_{J(My)} 
              \nonumber \\
        &=& -  \half \sq{{\rm Vol}(S^3)} \sum_{N,S} 
          C^{J+y m}_{J n, \half s} C^{J-y \prm}_{J \prn, \half \prs} 
           (\pd^{\bmu} Y_{J N})  Y_{\half S} ,
\eea 
where we have used the equation $x_{\blam}(\tau^{\blam})_{s\prs}
=r\sq{\fr{{\rm Vol}(S^3)}{2}} Y_{\half S}$. 
Using this expression and the identity  
$\pd_{\bmu}A\pd^{\bmu}B = \half \{ \Box (AB)-(\Box A)B-A\Box B \}$, 
where $\Box= \pd_{\bmu}\pd^{\bmu}$, we obtain 
\bba
   && \fr{r^2 {\rm Vol}(S^3)}{4}  \sum_{N_1,S} 
        C^{J_1+y_1 m_1}_{J_1 n_1, \half s}  
        C^{J_1-y_1 \prm_1}_{J_1 \prn_1, \half \prs} 
       \int_{S^3} d\Om_3 \Bigl\{  \Box (Y^*_{JM}Y_{J_1N_1}) 
          \nonumber      \\
   &&\qquad\qquad\qquad 
           -(\Box Y^*_{JM}) Y_{J_1N_1} 
                  -Y^*_{JM}\Box Y_{J_1N_1} \Bigr\} Y_{\half S}Y_{J_2M_2} .
\eea
This quantity can be evaluated using  the eigenequation for scalar 
harmonics, $\Box Y_{JM}=\fr{1}{r^2}\hnabla^2 Y_{JM}=-\fr{1}{r^2}2J(2J+2) Y_{JM}$, 
and the product expansion (\ref{prod-ss}) given in the next section. 
Furthermore, noting Eq.(\ref{transv-v}), we see that only 
the first term in the braces gives the non-vanishig quantity    
\bb 
   -\fr{1}{4} \sum_K 2K(2K+2) 
       \sum_{N_1,S,T} C^{J_1+y_1 m_1}_{J_1 n_1, \half s} 
        C^{J_1-y_1 \prm_1}_{J_1 \prn_1, \half \prs} 
        \C^{JM}_{J_1N_1,KT} \eps_S \C^{J_2M_2}_{KT,\half -S}. 
\ee
Using the formula (\ref{CCC}) relating three Clebsch-Gordan coefficients  
for the left and the right $SU(2)$ parts separately, we finally obtain  
\bba
  && \G^{JM}_{J_1(M_1y_1);J_2 M_2} 
           \nonumber \\
  && = -\fr{1}{2\sq{2}} \sq{ \fr{2J_1(2J_1+1)(2J_1+2)(2J_2+1)}{2J+1} } 
           \sum_{K=J_2 \pm \half} 2K(2K+1)(2K+2) 
              \nonumber   \\ 
  && \quad \times 
      \left\{ \begin{array}{ccc}
              J   &   J_1   & K \\
            \half &   J_2   & J_1+\half 
            \end{array} \right\}       
     \left\{ \begin{array}{ccc}
              J   &  J_1    & K \\
            \half &  J_2    & J_1-\half 
            \end{array} \right\}              
      C^{Jm}_{J_1+y_1 m_1,J_2 m_2} 
      C^{J\prm}_{J_1-y_1 \prm_1,J_2 \prm_2} . 
           \nonumber \\
\eea
{}From the properties of the standard Clebsch-Gordan coefficients, we find that 
this coefficient vanishes unless the triangular conditions
\bb
    |J_1 -J_2|+\half \leq  J  \leq J_1 +J_2 -\half , 
\ee
with half-integer $J+J_1+J_2$, and the requirement $M=M_1+M_2$ are satisfied.
Also, we find that these coefficients satisfy the relations
\bb
    \G^{JM}_{J_1(M_1y_1);J_2M_2}= -\G^{J-M}_{J_1(-M_1y_1);J_2-M_2}.
\ee

\paragraph{$\H$ coefficients}
The coefficient constructed from scalar, vector and tensor harmonics is 
\bba
    \H^{JM}_{J_1(M_1x_1);J_2(M_2y_2)} 
      &=& \sq{{\rm Vol}(S^3)} 
           \int_{S^3} d\Om_3 Y^*_{JM}Y^{ij}_{J_1(M_1x_1)}\hnabla_i Y_{j J_2(M_2y_2)} 
             \label{H-coeff} \\ 
      &=& -r^4 \sq{{\rm Vol}(S^3)}  
         \int_{S^3} d\Om_3 (\pd_{\bmu}Y^*_{JM}) 
          Y^{\bmu\bnu}_{J_1(M_1x_1)}Y_{\bnu J_2(M_2y_2)}.
             \nonumber
\eea
As in the case of the $\G$ coefficients, using $ x_{\bmu}Y^{\bmu\bnu}_{J(Mx)}=0$, 
we obtain the expression
\bba
    Y^{\bmu\bnu}_{J(Mx)}&=&-x_{\blam} \pd^{\bmu}Y^{\blam\bnu}_{J(Mx)} 
               \\ 
      &=& - \fr{1}{4r} \sq{\fr{3{\rm Vol}(S^3)}{2}}\sum_{N,T,U,S}
         C^{J+x m}_{J n,1t} C^{J-x \prm}_{J \prn, 1\prt}                        
         \C^{1 T}_{\half U,\half S}  
       (\pd^{\bmu} Y_{JN}) Y_{\half U} (\tau^{\bnu})_{s\prs} , 
          \nonumber 
\eea
where Eq.(\ref{tau3}) has been used. Substituting this expression into the definition, we obtain 
\bba
   && r^2 \fr{\sq{3}}{4} {\rm Vol}(S^3) \sum_{N_1,T_1}\sum_{N_2,S_2}\sum_U 
    C^{J_1+x_1 m_1}_{J_1 n_1, 1 t_1} C^{J_1-x_1 \prm_1}_{J_1 \prn_1, 1 \prt_1} 
    C^{J_2+y_2 m_2}_{J_2 n_2, \half s_2} C^{J_2-y_2 \prm_2}_{J_2 \prn_2, \half \prs_2} 
    \C^{\half U}_{1 T_1,\half S_2} 
         \nonumber   \\ 
   && \times 
     \int d\Om_3 (\pd_{\bmu} Y^*_{JM}) (\pd^{\bmu} Y_{J_1 N_1}) Y_{J_2 N_2} Y_{\half U} .
\eea  
The integral here is evaluated as in the computation of $\G$, and the product of 
Clebsch-Gordan coefficients are simplified using the formula (\ref{CCCCC}) 
for the left and the right $SU(2)$ parts separately and the formula 
for the 6j-symbols (\ref{six-j}).  We then obtain the following form:
\bba
  && \H^{JM}_{J_1(M_1x_1);J_2(M_2y_2)} 
           \nonumber \\
  && = -\fr{3}{2\sq{2}}\sq{ \fr{(2J_1-1)(2J_1+1)(2J_1+3)2J_2(2J_2+1)(2J_2+2)}{2J+1} } 
            \nonumber \\ 
  && \quad \times  
       \sum_{K=J_2 \pm \half} 2K(2K+1)(2K+2) 
      \left\{ \begin{array}{ccc}
              K   &   1    & J_2+y_2 \\
            \half &  J_2   & \half 
            \end{array} \right\}       
            \nonumber  \\ 
  && \quad \times 
     \left\{ \begin{array}{ccc}
              K   &  1   & J_2-y_2 \\
            \half & J_2  & \half 
            \end{array} \right\}              
     \left\{ \begin{array}{ccc}
              J  & J_1+x_1  & J_2+y_2 \\
              1  &    K     & J_1 
            \end{array} \right\}         
     \left\{ \begin{array}{ccc}
              J & J_1-x_1  & J_2-y_2 \\ 
              1 &    K     & J_1 
            \end{array} \right\}            
                \nonumber \\        
   && \quad \times  
      C^{Jm}_{J_1+x_1 m_1,J_2+y_2 m_2} 
      C^{J\prm}_{J_1-x_1 \prm_1,J_2-y_2 \prm_2}.
\eea
This vanishes unless the triangular conditions 
\bba
    |J_1 -J_2|+\half \leq &J& \leq J_1 +J_2 -\fr{3}{2} 
                    \qquad \hbox{for $x_1 = 2y_2$}, 
               \nonumber  \\
    |J_1 -J_2|+\fr{3}{2} \leq &J& \leq J_1 +J_2 -\half  
                        \qquad \hbox{for $x_1 \neq 2y_2$}, 
\eea
with half-integer $J+J_1+J_2$, and the requirement $M=M_1+M_2$ are satisfied. 
Also, the relations
\bb
    \H^{JM}_{J_1(M_1x_1);J_2(M_2y_2)}= -\H^{J-M}_{J_1(-M_1x_1);J_2(-M_2y_2)} 
\ee
are satisfied. 
   
\section{Product Expansions and Crossing Relations}
\setcounter{equation}{0}
\noindent

\paragraph{Product expansions}
The ${\rm ST}^2$ tensor harmonics satisfy the following product expansions:
\bba
   && Y_{J_1M_1}Y_{J_2M_2} = \fr{1}{\sq{{\rm Vol}(S^3)} }
       \sum_{J \geq 0}\sum_M \C^{JM}_{J_1M_1, J_2M_2}Y_{JM}, 
               \label{prod-ss}     \\
   &&  Y^i_{J_1(M_1y_1)}Y_{i J_2(M_2y_2)} = \fr{1}{\sq{{\rm Vol}(S^3)}}
          \sum_{J \geq 0}\sum_M \D^{JM}_{J_1(M_1y_1), J_2(M_2y_2)}Y_{JM}, 
              \label{prod-vv}      \\ 
   &&  Y^{ij}_{J_1(M_1x_1)}Y_{ij J_2(M_2x_2)} = \fr{1}{\sq{{\rm Vol}(S^3)}}
          \sum_{J \geq 0}\sum_M \E^{JM}_{J_1(M_1x_1), J_2(M_2x_2)}Y_{JM}.                 
               \label{prod-tt} 
\eea
The products here are taken at the same points. 
These equations are identified to the definitions of 
the $\C$, $\D$ and $\E$ coefficients.

  More useful product expansions are 
\bba
   && Y^*_{J_1M_1}Y^i_{J_2(M_2y_2)} 
          \nonumber  \\ 
   && = - \fr{1}{\sq{{\rm Vol}(S^3)}} \sum_{J \geq \half} \sum_{M,y}  
         \eps_M \D^{J_1M_1}_{J_2(M_2y_2),J(-My)} Y^i_{J(My)} 
            \nonumber    \\ 
   &&\quad  + \fr{1}{\sq{{\rm Vol}(S^3)}} 
         \sum_{J \geq \half} \sum_M  \fr{1}{2J(2J+2)} 
          \eps_M \G^{J_1M_1}_{J_2(M_2y_2);J-M}  \hnabla^i Y_{JM}     
             \label{prod-sv}
\eea 
and 
\bba
  && Y^*_{J_1M_1}Y^{ij}_{J_2(M_2x_2)} 
           \nonumber \\ 
  && = \fr{1}{\sq{{\rm Vol}(S^3)}} \sum_{J \geq 1} \sum_{M,x} 
         \eps_M \E^{J_1M_1}_{J_2(M_2x_2),J(-Mx)} Y^{ij}_{J(Mx)} 
            \nonumber    \\ 
  &&\quad  - \fr{1}{\sq{{\rm Vol}(S^3)}} 
          \sum_{J \geq 1} \sum_{M,y}  \fr{2}{(2J-1)(2J+3)} 
          \eps_M \H^{J_1M_1}_{J_2(M_2x_2);J(-My)}  \hnabla^{(i} Y^{j)}_{J(My)}  
             \nonumber \\ 
  &&\quad  +  \fr{1}{\sq{{\rm Vol}(S^3)}} 
        \sum_{J \geq 1} \sum_M \fr{3}{2} \fr{1}{(2J-1)2J(2J+2)(2J+3)}
                    \nonumber \\
   && \qquad\qquad\qquad \times
         \eps_M \I^{J_1M_1}_{J_2(M_2x_2);J-M}  
         \left( \hnabla^{(i} \hnabla^{j)} -\fr{1}{3}\hgm^{ij}\hnabla^2 
                                                \right) Y_{JM} ,    
            \label{prod-st}
\eea
where the smallest values of the sums of $J$ are determined by Eqs.(\ref{special-s}) 
and (\ref{special-v}). The new coefficient is defined by 
\bb
   \I^{JM}_{J_1(M_1x_1);J_2M_2} 
   = \sq{{\rm Vol}(S^3)} \int_{S^3} d\Om_3 
     Y^*_{JM} Y^{ij}_{J_1(M_1x_1)}\hnabla_i \hnabla_j Y_{J_2M_2} .  
\ee   
Using Eq.(\ref{special-s}), we see that this coefficient vanishes for $J=\half$: 
\bb
   \I^{\half M}_{J_1(M_1x_1);J_2M_2}=0. 
      \label{I1/2}
\ee
This fact is important when we construct the conformal algebra in the traceless mode sector.

   Furthermore, we consider 
\bba
  && Y^*_{J_1M_1}\hnabla^{(i} Y^{j)}_{J_2(M_2y_2)} 
           \nonumber \\
  && = \fr{1}{\sq{{\rm Vol}(S^3)}} \sum_{J \geq 1} \sum_{M,x}  
         \eps_M \H^{J_1M_1}_{J(-Mx);J_2(M_2y_2)} Y^{ij}_{J(Mx)} 
            \nonumber    \\ 
  &&\quad  - \fr{1}{\sq{{\rm Vol}(S^3)}} 
          \sum_{J \geq 1} \sum_{M,y}  \fr{1}{(2J-1)(2J+3)} 
            \biggl[ \half \Bigl\{ 2J(2J+2)
                     \nonumber  \\ 
  && \qquad\quad  
          -2J_1(2J_1+2)+2J_2(2J_2+2)-6 \Bigr\}
           \eps_M \D^{J_1M_1}_{J_2(M_2y_2),J(-My)}   
             \nonumber \\ 
  && \qquad\qquad\qquad\qquad\qquad
          + \eps_M \tD^{J_1M_1}_{J_2(M_2y_2),J(-My)} 
                 \biggr]  \hnabla^{(i} Y^{j)}_{J(My)}
            \nonumber  \\     
  &&\quad  + \fr{1}{\sq{{\rm Vol}(S^3)}} 
       \sum_{J \geq 1} \sum_M \fr{3}{4}\fr{1}{(2J-1)2J(2J+2)(2J+3)}
            \nonumber  \\
  && \qquad \times
          \Bigl\{ 2J(2J+2)-2J_1(2J_1+2)+2J_2(2J_2+2)-3 \Bigr\} 
             \nonumber \\
  && \qquad \times           
        \eps_M \G^{J_1M_1}_{J_2(M_2y_2);J-M}  
         \left( \hnabla^{(i} \hnabla^{j)} -\fr{1}{3}\hgm^{ij}\hnabla^2 
                                                \right) Y_{JM} ,   
               \label{prod-sdv}
\eea
where
\bb
    \tD^{JM}_{J_1(M_1y_1),J_2(M_2y_2)} 
    = \sq{{\rm Vol}(S^3)} \int_{S^3} d\Om_3 (\hnabla_i \hnabla_j Y^*_{JM} )
         Y^i_{J_1(M_1y_1)} Y^j_{J_2(M_1y_2)}.   
\ee
The $\tD$ coefficients have the same indices as the $\D$ coefficients. 
For $J=\half$, using Eq.(\ref{special-s}), we find that 
\bb
    \tD^{\half M}_{J_1(M_1y_1),J_2(M_2y_2)} 
    = -\D^{\half M}_{J_1(M_1y_1),J_2(M_2y_2)} .    
          \label{tD1/2}
\ee
This equation is used below.

\paragraph{Crossing relations}
Let us consider the following integral of the product of four scalar harmonics: 
\bb
   \int_{S^3} d\Om_3 Y^*_{J_1M_1}Y_{J_2M_2}Y^*_{J_3M_3}Y_{J_4M_4}. 
         \label{cross-ss} 
\ee
Using the product expansion (\ref{prod-ss}), we obtain the following 
crossing relation:
\bb
    \sum_{J \geq 0} \sum_M \eps_M \C^{J_1M_1}_{J_2M_2, J-M} 
                               \C^{J_3M_3}_{JM,J_4M_4} 
    = \sum_{J \geq 0} \sum_M \eps_M \C^{J_1M_1}_{J_4M_4, J-M} 
                               \C^{J_3M_3}_{JM,J_2M_2} .                          
           \label{cross-cc}
\ee  
The case $J_1=J_3=\half$ is used to derive the conformal algebra given in Sect. 4, 
and, also, the case $J_1=\half$ and $J_3=J+J_4$ is used to derive the creation 
operator that commutes with the charge $Q_M$, given in Sect. 5.

  As a variant of this crossing relation, we consider the relation obtained from the 
integral in which $Y_{J_2M_2}$ and $Y_{J_4M_4}$ are replaced with other harmonics.  
Changing to vector harmonics and using the product 
expansion (\ref{prod-sv}), we obtain
\bba
  && \sum_{J \geq \half} \sum_{M,y} \eps_M 
      \D^{J_1M_1}_{J_2(M_2y_2), J(-My)} \D^{J_3M_3}_{J(My),J_4(M_4y_4)} 
             \nonumber \\ 
  && -\sum_{J \geq \half} \sum_M \fr{1}{2J(2J+2)} \eps_M 
      \G^{J_1M_1}_{J_2(M_2y_2); J-M} \G^{J_3M_3}_{J_4(M_4y_4);JM} 
             \nonumber  \\ 
  && =[J_2 (M_2y_2) \leftrightarrow J_4 (M_4y_4)], 
\eea 
and, also, changing to tensor harmonics and using (\ref{prod-st}), we obtain 
\bba
  && \sum_{J \geq 1} \sum_{M,x}  \eps_M 
      \E^{J_1M_1}_{J_2(M_2x_2), J(-Mx)} \E^{J_3M_3}_{J(Mx),J_4(M_4x_4)} 
             \nonumber \\ 
  && - \sum_{J \geq 1} \sum_{M,y} \fr{2}{(2J-1)(2J+3)} \eps_M 
      \H^{J_1M_1}_{J_2(M_2x_2); J(-My)} \H^{J_3M_3}_{J_4(M_4x_4);J(My)} 
             \nonumber  \\ 
  && +\sum_{J \geq 1}\sum_M \fr{3}{2}\fr{1}{(2J-1)2J(2J+2)(2J+3)} \eps_M 
       \I^{J_1M_1}_{J_2(M_2x_2); J-M)} \I^{J_3M_3}_{J_4(M_4x_4);JM} 
             \nonumber  \\   
  && =[J_2 (M_2x_2) \leftrightarrow J_4 (M_4x_4)] .
               \label{cross-ee} 
\eea

  Next, consider the integral
\bb
     \int_{S^3} d\Om_3 Y^*_{J_1M_1}Y^{ij}_{J_2(M_2x_2)}
                   Y^*_{J_3M_3} \hnabla_{(i}Y_{j)J_4(M_4y_4)}.
\ee
For simplicity, we consider the case $J_1=J_3=\half$ used in the text. 
Then, using the products (\ref{prod-st}) and (\ref{prod-sdv}), we obtain 
\bba
   && \sum_{J=J_4}\sum_{M,x}  \eps_M 
          \E^{\half M_1}_{J_2(M_2x_2), J(-Mx)} \H^{\half M_3}_{J(Mx);J_4(M_4y_4)} 
             \nonumber \\ 
   && - \sum_{J=J_2} \sum_{M,y} \half \fr{1}{(2J-1)(2J+3)} 
        \Bigl\{ 2J(2J+2)+2J_4(2J_4+2) -11 \Bigr\} 
             \nonumber \\ 
   &&\qquad\qquad\qquad\qquad\qquad \times
    \eps_M \H^{\half M_1}_{J_2(M_2x_2); J(-My)} \D^{\half M_3}_{J(My),J_4(M_4y_4)} 
             \nonumber  \\ 
   && = \sum_{J=J_4} \sum_{M,x} \eps_M 
          \H^{\half M_1}_{J(-Mx); J_4(M_4y_4)} \E^{\half M_3}_{J(Mx),J_2(M_2x_2)} 
             \nonumber \\ 
   &&\quad - \sum_{J=J_2} \sum_{M,y} \half \fr{1}{(2J-1)(2J+3)} 
        \Bigl\{ 2J(2J+2)+2J_4(2J_4+2) -11 \Bigr\} 
             \nonumber \\ 
   &&\qquad\qquad\qquad\qquad   \times     
  \eps_M \D^{\half M_1}_{J_4(M_4y_4), J(-My)} \H^{\half M_3}_{J_2(M_2x_2);J(My)},  
            \label{cross-eh}  
\eea
where we have used Eqs. (\ref{I1/2}) and (\ref{tD1/2}). The values of $J$ are now 
fixed, because $\H^{\half M}_{J_1 (M_1 x_1); J_2 (M_2 y_2)} \propto \dl_{J_1 J_2}$.

  Then, consider the integral
\bba
   &&  \int_{S^3} d\Om_3 \hnabla_i \left( 
                  Y^*_{\half M_1} Y^{ij}_{J_2(M_2x_2)} \right)
                   Y^*_{\half M_3} Y_{j J_4(M_4y_4)}
            \nonumber  \\ 
   && = -\int_{S^3} d\Om_3  Y_{\half M_1} Y^{ij}_{J_2(M_2x_2)} 
                   (\hnabla_i Y^*_{\half M_3}) Y_{j J_4(M_4y_4)} 
             \nonumber \\ 
   && \quad 
        - \int_{S^3} d\Om_3 Y^*_{\half M_1} Y^{ij}_{J_2(M_2x_2)}
                   Y^*_{\half M_3} \hnabla_i Y_{j J_4(M_4y_4)} .  
\eea
In the integral on the l.h.s. here, the product of $1$ and $2$ and of $3$ and $4$ 
are evaluated first, and in the two integrals on the r.h.s., the products of $1$ and $4$ 
and of $2$ and $3$ are evaluated first.
Then, we obtain 
\bba
   && \sum_{J=J_2}\sum_{M,y}  \eps_M 
      \H^{\half M_1}_{J_2(M_2x_2);J(-My)} \D^{\half M_3}_{J(My),J_4 (M_4y_4)} 
           \nonumber \\ 
   && = \sum_{J=J_2} \sum_{M,y} \fr{1}{(2J-1)(2J+3)} 
         \left[ \half \left\{ 2J(2J+2)+2J_4(2J_4+2)-11 \right\} -1 \right] 
            \nonumber \\ 
   && \qquad\qquad\qquad\qquad\qquad \times
    \eps_M \D^{\half M_1}_{J_4(M_4y_4),J(-My)} \H^{\half M_3}_{J_2(M_2x_2);J (My)}                  
             \nonumber \\
   && \quad
      - \sum_{J=J_4}\sum_{M,x} \eps_M 
         \H^{\half M_1}_{J(-Mx);J_4(M_4y_4)} \E^{\half M_3}_{J(Mx),J_2 (M_2x_2)} . 
             \label{cross-hd}
\eea

\section{Some Important Fomulae for Clebsch-Gordan Coefficients 
and Wigner $D$ functions~\cite{vmk}}
\setcounter{equation}{0}
\noindent 

  The standard Clebsch-Gordan coefficient $C^{c\gm}_{a\a,b\b}$ vanishes unless 
the triangular condition $|a-b| \leq c \leq a+b$ and the 
requirement $\a+\b=\gm$ are satisfied, where $a$, $b$ and $c$ are integer or half-integer, 
non-negative numbers, and $a+b+c$, $a+\a$, $b+\b$, $c+\gm$ are integer, non-negative 
numbers. The coefficients satisfy the equations  
\bb
       C^{c\gm}_{a\a,b\b}=(-1)^{a+b-c}C^{c-\gm}_{a-\a,b-\b}
       =(-1)^{a+b-c}C^{c\gm}_{b\b,a\a}
       =(-1)^{b+\b}\sq{ \fr{2c+1}{2a+1} }C^{a-\a}_{c-\gm,b\b}.
               \label{property-C}
\ee

  The formulae for the products of the Clebsch-Gordan coefficients are as follows: 
\bb
    \sum_{\a,\b}C^{c\gm}_{a\a,b\b}C^{c^{\pp}\gm^{\pp}}_{a\a,b\b}
    =\dl_{cc^{\pp}}\dl_{\gm\gm^{\pp}}, 
           \label{CC}
\ee
\bba
 && \sum_{\a,\b,\dl} (-1)^{a-\a} C^{c\gm}_{b\b,a\a}C^{e\eps}_{b\b,d\dl}
     C^{f\vphi}_{d\dl,a-\a} 
            \nonumber \\
 && =(-1)^{a+b+e+f} \sq{(2c+1)(2f+1)} 
            C^{e\eps}_{c\gm, f\vphi} 
        \left\{ \begin{array}{ccc}
                  a & b & c \\ 
                  e & f & d 
                \end{array}  \right\}  ,
            \label{CCC}
\eea
\bba
 && \sum_{\psi,\kappa,\rho,\s,\tau} (-1)^{\psi+\kappa+\rho+\s+\tau} 
     C^{a\a}_{p\psi,q\kappa}C^{b\b}_{q\kappa,r\rho}
     C^{c\gm}_{r\rho,s\s} C^{d\dl}_{s\s,t\tau} C^{e\veps}_{t\tau,p-\psi} 
            \nonumber \\
 && = (-1)^{-a-b-2c-2p-2r-t+\a+\dl} \sq{(2a+1)(2d+1)} 
             \nonumber \\
 && \quad \times  
     \sum_{x,y}\sum_{\xi,\eta}(-1)^{\xi+\eta}(2x+1)(2y+1) 
      C^{b\b}_{a\a,x\xi} C^{e\veps}_{x\xi,y\eta} C^{c-\gm}_{y\eta,d-\dl}
            \nonumber \\
 && \quad \times   
     \left\{  \begin{array}{ccc}
               a & b & x \\ 
               r & p & q 
               \end{array} \right\}   
      \left\{  \begin{array}{ccc}
               x & e & y \\ 
               t & r & p 
               \end{array} \right\}     
       \left\{  \begin{array}{ccc}
               y & c & d \\ 
               s & t & r 
               \end{array} \right\}  .
             \label{CCCCC}
\eea

  Some important properties of the Wigner $D$ functions are given below: 
\bb
    D^{J*}_{m\prm} = (-1)^{m-\prm} D^J_{-m-\prm} , 
          \label{D1}
\ee
\bb
   D^{J_1}_{m_1\prm_1} D^{J_2}_{m_2\prm_2}  
   =\sum_{J=|J_1-J_2|}^{J_1+J_2} \sum_{m,\prm} 
      C^{Jm}_{J_1m_1,J_2m_2} C^{J\prm}_{J_1\prm_1,J_2\prm_2} 
      D^J_{m\prm} ,
           \label{D2}
\ee
\bb
   \sum_{m_1,\prm_1}\sum_{m_2,\prm_2} C^{Jm}_{J_1m_1,J_2m_2} 
          C^{J^{\pp}\prm}_{J_1\prm_1,J_2\prm_2}  
            D^{J_1}_{m_1\prm_1}  
            D^{J_2}_{m_2\prm_2}  
       =\dl_{JJ^{\pp}} \{ J_1J_2J \} D^J_{m\prm} ,
           \label{D3}
\ee
\bba
  && \sum_{m_1,\prm_1}\sum_{m_2,\prm_2}\sum_{m_3,\prm_3} 
          C^{Jm}_{Kn,J_3m_3} 
          C^{Kn}_{J_1m_1,J_2m_2} 
          C^{J^{\pp}\prm}_{K^{\pp}\prn,J_3\prm_3}
          C^{K^{\pp}\prn}_{J_1\prm_1,J_2\prm_2} 
              \nonumber  \\ 
   &&  \times 
            D^{J_1}_{m_1\prm_1}  
            D^{J_2}_{m_2\prm_2}  
            D^{J_3}_{m_3\prm_3}  
       =\dl_{JJ^{\pp}}\dl_{KK^{\pp}} 
         \{ J_1 J_2 K \} \{ K J_3 J\} D^J_{m\prm} ,
           \label{D4}
\eea
\bba
     \hnabla^2 D^J_{m\prm} 
     &=&4\left\{ \pd^2_{\b} +\cot \b \pd_{\b} +\fr{1}{\sin^2 \b} 
                \left( \pd^2_{\a} -2\cos \b \pd_{\a}\pd_{\gm}+\pd^2_{\gm} \right)
          \right\} D^J_{m\prm} 
            \nonumber \\ 
     &=& -4J(J+1)D^J_{m\prm},
           \label{D5}
\eea
\bb
    \int_{S^3} d\Om_3 D^{J_1*}_{m_1\prm_1}D^{J_2}_{m_2\prm_2} 
    =\fr{{\rm Vol}(S^3)}{2J_1+1} \dl_{J_1J_2}\dl_{m_1m_2}\dl_{\prm_1\prm_2},
            \label{D6}
\ee
\bb
    \sum^J_{\prm=-J} D^{J*}_{m_1\prm} D^J_{m_2\prm}  
    = \dl_{m_1m_2}. 
          \label{D7}
\ee
Here, the arguments of the $D$ functions 
are all $D^J_{m\prm}=D^J_{m\prm}(\a,\b,\gm)$. 
The quantity $\{ J_1 J_2 J_3 \}$ is unity if $J_1+J_2+J_3$ is an integer 
and $|J_1-J_2| \leq J_3 \leq J_1+J_2$, while it vanishes otherwise. 
$\{ J_1 J_2 J_3 \}$ is invariant with respect to permutations 
of $J_1$, $J_2$ and $J_3$.

  The formula for the products of $6j$-symbols is
\bb
    \sum_x (-1)^{p+q+x}(2x+1)
      \left\{  \begin{array}{ccc}
               a & b & x \\ 
               c & d & p 
               \end{array} \right\}   
      \left\{  \begin{array}{ccc}
               a & b & x \\ 
               d & c & q 
               \end{array} \right\}     
      = \left\{  \begin{array}{ccc}
               a & c & q \\ 
               b & d & p 
               \end{array} \right\} . 
            \label{six-j}
\ee

\section{$R_{MN}$ for the conformal mode and the traceless mode sectors}
\setcounter{equation}{0}
\noindent

  Using the parametrization (\ref{parametrization}), the charges $R_{MN}$ for the conformal 
mode and the traceless mode sectors are obtained as 
\bba
  R_{11} &=& \sum_{J > 0}\sum_M (m+\prm) \left( 
             a^{\dag}_{JM} a_{JM} -b^{\dag}_{JM} b_{JM} \right)
                \nonumber \\
     &&  +\sum_{J \geq 1}\sum_{M,x} (m+\prm) \left( 
             c^{\dag}_{J(Mx)}c_{J(Mx)}- d^{\dag}_{J(Mx)}d_{J(Mx)} \right) 
                    \\
     && -\sum_{J \geq 1}\sum_{M,y} (m+\prm)  
             e^{\dag}_{J(My)}e_{J(My)} , 
                \nonumber   \\
  R_{22} &=& \sum_{J > 0}\sum_M (m-\prm)  \left( 
             a^{\dag}_{JM} a_{JM} - b^{\dag}_{JM} b_{JM} \right) 
                  \nonumber \\
     && +\sum_{J \geq 1}\sum_{M,x} (m-\prm) \left( 
             c^{\dag}_{J(Mx)}c_{J(Mx)}- d^{\dag}_{J(Mx)}d_{J(Mx)} \right) 
                      \\
     && -\sum_{J \geq 1}\sum_{M,y} (m-\prm)  
             e^{\dag}_{J(My)}e_{J(My)} ,   
               \nonumber  \\
  R_{21} &=& \half \sum_{J > 0}\sum_M \sq{(2J+2-2\prm)(2J+2\prm)} \left( 
             a^{\dag}_{JM}a_{J\bM} - b^{\dag}_{JM}b_{J\bM} \right) 
               \nonumber  \\   
     && + \half \sum_{J \geq 1}\sum_M \sq{(2J-2\prm)(2J-2+2\prm)} 
               \nonumber  \\ 
     && \qquad\qquad\qquad \times 
        \left( c^{\dag}_{J(M1)}c_{J(\bM1)}- d^{\dag}_{J(M1)} d_{J(\bM1)} \right) 
                 \nonumber \\
     && - \half \sum_{J \geq 1}\sum_M \sq{(2J+1-2\prm)(2J-1+2\prm)} 
             e^{\dag}_{J(M\half)}e_{J(\bM\half)} 
                     \\ 
     && + \half \sum_{J \geq 1}\sum_M \sq{(2J+4-2\prm)(2J+2+2\prm)} 
               \nonumber  \\ 
     && \qquad\qquad\qquad \times 
     \left( c^{\dag}_{J(M-1)}c_{J(\bM-1)}- d^{\dag}_{J(M-1)}d_{J(\bM-1)} \right) 
                 \nonumber \\ 
     & & - \half \sum_{J \geq 1}\sum_M \sq{(2J+3-2\prm)(2J+1+2\prm)}  
             e^{\dag}_{J(M-\half)}e_{J(\bM-\half)}, 
                 \nonumber \\ 
  R_{31} &=& \half \sum_{J > 0}\sum_M \sq{(2J+2-2m)(2J+2m)} \left(  
             a^{\dag}_{JM} a_{J\uM} - b^{\dag}_{JM} b_{J\uM} \right) 
                 \nonumber \\   
     && + \half \sum_{J \geq 1}\sum_{M} \sq{(2J+4-2m)(2J+2+2m)} 
               \nonumber  \\ 
     && \qquad\qquad\qquad \times 
       \left( c^{\dag}_{J(M1)}c_{J(\uM1)}- d^{\dag}_{J(M1)}d_{J(\uM1)} \right) 
                 \nonumber \\
     && - \half \sum_{J \geq 1}\sum_M \sq{(2J+3-2m)(2J+1+2m)}  
             e^{\dag}_{J(M\half)}e_{J(\uM\half)} 
                       \\ 
     && + \half \sum_{J \geq 1}\sum_{M} \sq{(2J-2m)(2J-2+2m)} 
               \nonumber  \\ 
     && \qquad\qquad\qquad \times 
       \left( c^{\dag}_{J(M-1)}c_{J(\uM-1)}- d^{\dag}_{J(M-1)}d_{J(\uM-1)} \right) 
                 \nonumber \\
     && - \half \sum_{J \geq 1}\sum_M \sq{(2J+1-2m)(2J-1+2m)}  
             e^{\dag}_{J(M-\half)}e_{J(\uM-\half)}, 
                \nonumber 
\eea
where $\bM=(m,\prm-1)$ and $\uM=(m-1,\prm)$.

\end{document}